\documentclass{emulateapj} 
\usepackage{natbib} 
\usepackage{graphicx}

\slugcomment{2010 ApJ 716, 634}% 

\begin{document} 
 
 
\shorttitle{FLAMINGOS Survey of the Serpens Cloud}
\shortauthors{Gorlova et al.} 

\title{FLAMINGOS Near Infra-Red Survey \\
 of The Serpens Cloud Main Core} 


\author{Nadya Gorlova$^{1,2,3}$,Aaron Steinhauer$^{3,4}$, Elizabeth Lada$^{1,3}$}
\affil{$^1$Department of Astronomy, University of Florida, Gainesville, FL 32611-2055}
\affil{$^2$Current address: Institute of Astronomy, KU Leuven, Celestijnenlaan 200D, 3001, Leuven, Belgium {\it nadya@ster.kuleuven.be}}
\affil{$^3$Visiting astronomer, Kitt Peak National Observatory}
\affil{$^4$Department of Physics and Astronomy, SUNY Geneseo, One College Circle, Geneseo, NY 14454}
%% identified by the \altaffilmark after each name.  Specify alternate 
 

\begin{abstract}

We obtained $JHK$ images and low-resolution $JH$ spectra
in the embedded young cluster in the Serpens cloud Main core
(also known as Serpens North).
We determined spectral types (SpT)
for 15 previously identified cluster members (for 5 of them 
for the first time),
1 new candidate, and 11 stars that appear to be field interlopers.
Extinction, for which we derived an analytical expression,
was obtained by taking SpT and near-IR excess
into account.
The location on the Hertzsprung-Russell
diagram indicates that we probed
a low-mass population of the cloud (0.05 -- 1.5 M$_{\sun}$),
including 1 -- 3 brown dwarfs.
We used our individually determined
photospheric parameters
to analyze the {\it ISO} and the {\it Spitzer} determined
spectral energy distribution (SED) classes. The latter were correlated
with the age and location of the sources in the cloud.
We find that most flat objects from our study (4 out of 5) have SEDs consistent
with reddened classical T Tau stars; however,
when comparing to the thick disk SEDs of
lower mass {\it M-type} objects,
we find that the flat ones show more excess,
perhaps indicating an earlier evolutionary stage.
We determined a median age for the cluster to
be 1 Myr for distance of 380 pc, and 3 Myr for
a less likely distance of 260 pc.
The core of the cluster is on average younger than the
rest of the cluster. We do not find objects with disks past 5 Myr.
We do find diskless, X-ray bright objects younger than 1 Myr,
as was also noted in the study of Winston et al (2009).
We find two groups of young objects associated
with dark filaments, indicating that star formation
was not always confined to the core.
 
\end{abstract}


\section{Introduction} 

Embedded young clusters are where the earliest stages of star formation
are taking place. As a result of the interplay of gravity,
turbulence, external pressure, and magnetic fields,
molecular clouds fragment into filaments and cores.
Then protostars with envelopes and disks develop,
and finally single and multiple stars with planets evolve.
Once the gas dissipates, stars leave their birthplaces
due to velocity dispersion and mutual interactions \citep{Bate03}.
The outcome can range from a moving group of a dozen
young stars to a rich super-cluster of a few thousand members
\citep{Lada03}.
 
When, where, and for how long the cloud fragmentation proceeds
is currently not fully understood.
Although evidence is accumulating in favor
of triggered sequential star formation \citep[e.g.,][]{Lee07},
small statistics and the large observational uncertainties of extinction-limited optical surveys
can not provide definitive answers to these questions.
An age spread of a few tenths of Myr is often observed in a given cluster,
while the use of different theoretical isochrones may
result in markedly different mean ages
\citep{Lada03, Hillenbrand05, Hillenbrand08}.
These results call for infra-red (IR) surveys,
that are also better at studying cool low-mass objects,
including brown dwarfs and free floating planets
\citep[e.g.,][]{Muench03, Luhman07, Andersen08}.
In 2003 the {\it Spitzer} Space Telescope was launched.
Its un-precedented sensitivity
and mapping capabilities allowed an extensive
study of the circumstellar (and circum-substellar) matter in cluster stars.
A large variation of disk properties in a given cluster
was found in these surveys \citep[e.g.,][]{Cieza07}.
A number of explanations have been proposed,
however, none is widely accepted yet:
(1) diskless stars are older than the ones with disks \citep{Bertout07};
(2) diskless stars are more massive since they disperse their disks faster \citep{Carpenter06},
e.g. due to photo-evaporation \citep{Hollenbach94};
(3) low-mass disks can also be evaporated by nearby O stars \citep{Balog06}; 
(4) in some systems giant planets may form early on and
sweep up the disk \citep{Quillen04};
(5) disks are truncated by close companions \citep{Bowman06},
or (6) during the close passage of another star \citep{Olczak06}.
Young clusters provide ideal places to test these hypotheses.

The first step toward deciphering the star formation history in the cloud
is to identify its population.
At the typical distances of a few hundred parsecs,
members are most efficiently identified through the mid-IR excess
from circumstellar material.
{\it Spitzer} investigated a number of molecular clouds
and uncovered many new clusters as well as distributed populations
of young stellar objects (YSOs) \citep{Allen07}.
The IR excess method, however, misses YSOs that have already dispersed
their circumstellar material. To obtain a total count
of the cloud population, complementary less reliable methods are being used,
such as association with nebulosities, X-rays, and variability surveys
that detect YSOs by the elevated magnetic activity.

Once cloud members are identified,
the next step is to characterize the age and the evolutionary state
of each one. This requires the determination of photospheric luminosity
and temperature for comparison with evolutionary tracks.
Obscuration by dust makes traditional optical techniques
(like multi-color photometry or spectroscopy) inefficient.
The mid-IR region, on the other hand, is not sensitive to the photospheric temperature,
falling on the Rayleigh-Jeans tail of the stellar flux.
Near-IR (NIR) photometry
would be optimal, however, it is not sufficient by itself.
The leverage in wavelength space is too small
for the accurate determination of the effective temperature and reddening.
For example, late-type YSOs can be mistaken with reddened 
early-type YSOs and with background evolved stars.
The latter is inevitable since molecular clouds
are usually confined to the Galactic plane \citep{Dame01}.
These obstacles can be overcome by obtaining NIR spectra.
We designed our Molecular Cloud Survey
to gather both deep NIR images of dark clouds and NIR spectra of
cloud sources, using the multi-object capabilities of the FLAMINGOS instrument
(Florida Multi-Object Imaging Near-Infrared Grism Observational
Spectrometer, \citet{Elston03}).
The Serpens cloud is one of the targets of this survey.

The cluster of protostars at the heart of the dark cloud
in Serpens Cauda has been known since the mid 70's.
It was extensively studied since then
at all wavelengths, and new concentrations of YSOs were traced
in the southern direction.
Young objects in all evolutionary stages were found throughout the cloud,
indicating a prolonged star formation history.
Despite the large body of research, a Hertzsprung-Russell (HR) diagram
with a representative number of sources did not exist until recently
\citep{Wilking08, Winston09}.
Earlier, \citet{Doppmann05} attempted to derive ages
for nine Class I and flat-spectrum Serpens protostars
using an alternative form of the HR diagram,
that employs spectroscopically-determined surface
gravity in place of luminosity.
They obtained an age spread $>$3 Myr,
too large on statistical and physical grounds
for a protostellar stage that is expected to last $\sim$0.1 Myr.
\citet{Giov98} and \citet{Kaas04} modeled the
K-band luminosity function around the Main core and
concluded that protostars represent a recent
``micro-burst'', while a more evolved distributed
bulk population originated 2--3 Myr ago, perhaps
in a similar burst.

It is difficult to correct luminosity functions
for the background population in this region
because of the highly non-uniform extinction
and proximity to the galactic plane ($b=5\degr$).
Recently an alternative spectroscopic approach
has been attempted \citep{Wilking08,Oliveira09,Winston09}.
Our study follows this approach with
the first NIR multi-object spectrograph.  
Our primary goal is to carry out spectral classification for
the accurate age determination.
We also want to understand the age spread, by
considering the spatial location
and the evolutionary status of individual objects,
based on the {\it Chandra} and {\it Spitzer} SED classification.
Spectroscopy also allows to identify interesting emission-line objects
and brown dwarf members (of which only one has been
closely scrutinized by \citet{Lodieu02}).

\section{The Serpens Cloud}\label{descrp1}

A chain of dark clouds in Serpens Cauda is part of the Aquila Rift,
a larger complex of molecular clouds that branches off the Milky Way
plane in Cygnus, runs through northern parts of Aquila and Serpens Cauda,
and disappears in Ophiuchus.
One of the prominent sources in the cloud is
an optically visible Herbig Ae/Be star
VV Ser ($A_{V}=3.0^{m}, M\sim 3M_{\sun}$ for $d\sim$400pc, \citet{Hillenbrand92}),
surrounded by a mid-IR nebula \citep{Pontoppidan07a, Pont07b}.
The region covered in our study lies inside a dark filamentary cloud
1$\degr$ north of VV Ser and is called the Main Serpens Core
(Figure \ref{dss_irac}).
It was first noticed due to the red reflection nebula,
the Serpens Reflection Nebula (SRN). 
\citet{Strom74, Strom76} conducted a NIR survey
of the cloud and identified the first
embedded population (now referred to as ``cluster A''),
including SVS20, a bright IR source
with a ring-like nebulosity, and SVS2, an illuminating source
of the SRN \citep{Worden74}.
An NH$_{3}$ map revealed an elongated cloud of dense gas
that stretches 10$\arcmin$ south and bends 4$\arcmin$ north-west
from the SRN \citep{Little80}, following contours of high extinction
\citep{Cambresy99, Harvey06}.
Sub-mm and mm surveys of the dust continuum emission
revealed a double core with clumps, most of which (but not all)
harbor mid-IR sources
\citep{Casali93, Davis99, Kaas04, Enoch07, Enoch08}.
Unlike the southern sub-core, the north-western one is devoid of bright NIR sources,
but teems with outflows powered by deeply embedded class 0 and I protostars
\citep{Bally83, Davis99, White95, Ziener99}.

Following the initial NIR studies of the region \citep{Strom76,Cohen79,Churchwell86},
\citet{Eiroa92} assembled the first representative list of
$\sim$50 potential cloud members. Membership criteria included
association with a nebulosity or mini-cluster,
NIR excess, emission lines, and the 3 $\mu$m ice absorption feature.
They estimated a reddening of $A_{V}=1-39$.
More members were identified in the subsequent NIR surveys
of \citet{Sogawa97}, \citet{Giov98}, and \citet{Kaas99}.
In the latter study 39 members without NIR excess were added
based on their $K$--band variability.
The low IR luminosity of the region implies
that it is devoid of O/B stars.
The nearest massive stars
are three late type B stars that illuminate reflection nebulosities at the
north-east rim of the cloud
(HD 170784, HD 170739, \& HD 170634, see Figure 1 in \citet{Zhang88a}).
Being detached from the main concentrations of protostars,
they are likely field stars passing through the cloud.
They have been used
for the estimation of the distance to the cloud (\S \ref{dist}).  

The Serpens cloud was surveyed with all three major IR space observatories:
{\it IRAS} \citep{Zhang88a, Zhang88b}, {\it ISO} \citep{Kaas04, Djupvik06}, and {\it Spitzer}.
Two {\it Spitzer} teams independently studied
the central $0.5\degr \times 0.5 \degr$ region:
the Legacy Cores to Disks (c2d) group \citep{Harvey06, Harvey07a, Harvey07b},
and the IRAC and MIPS instrument team \citep{Winston07}.
The c2d group in addition extended coverage  $2\degr$ south,
following the $A_{V}= 5$ contours of \citet{Cambresy99}.
The {\it Spitzer} studies provided the most comprehensive
census of the YSO population in the cloud.
Sources were assigned a spectral energy distribution (SED) class
and their spatial distribution was examined.
Protostars (class 0 \& I sources)
showed a strong clustering around two main dusty ridges --
the Main core with its cluster A, and another filament
42 $\arcmin$ south with a new cluster dubbed ``cluster B'' \citep{Harvey06, Enoch07}.
The latter is also known under the Ser/G3-G6 name \citep{Djupvik06},
after 4 classical T Tau stars (CTTS) found by \citet{Cohen79}.  
Soon after, \citet{Gutermuth08} discovered yet another
pocket of protostars, ``Serpens South'',
located $3\degr$ south of the Main core.
On the other hand, class II and III sources were found to be nearly uniformly distributed
throughout the cloud as defined by $A_{V}=4-5^{m}$ contours.
The resolution of the extinction maps used in these {\it Spitzer}
studies and of the mm images, however, was not sufficient to
explore association with dusty filaments in detail.
Our deep NIR FLAMINGOS image of the region (Figure \ref{flam_big}) makes
this investigation possible (\S \ref{sfh}). 

Due to the high sensitivity of {\it Spitzer}
and the negligible extinction in the mid-IR, the cloud
population constitutes only a miniscule fraction of all the sources
in the field of view (FOV).
The vast majority are extragalactic objects (such as star-forming galaxies
and active galactic nuclei) and field stars.
Extragalactic point sources are eliminated based on the faintness
at 2--6 $\mu$m and the large excess at 5.8 and 8 $\mu$m due PAHs,
combined with the absence of excess at shorter wavelengths.
\citet{Harvey07b} undertook a more conservative selection of YSOs
against AGB stars and galaxies than \citet{Winston07},
by selecting brighter sources and with a more robust 2MASS/IRAC excess.
This approach helps to remove spurious detections
in the case of close companions
or nebulosity, as shown by cyan squares in Figure \ref{flam_big},
however, it can miss sources where the excess starts beyond 5 $\mu$m, 
such as W68, W109, and W181.

Magnitude cuts of the IR surveys limit the ability to detect very low-mass YSOs,
class III are indistinguishable from field stars,
and weak/transition disks are missed due to loss of sensitivity in the {\it Spitzer}
IRAC 6 \& 8 $\mu$m channels. 
In the latter two cases X-ray observations are used to discriminate between
X-ray quiet slowly rotating field giants / Main Sequence stars and X-ray active YSOs.
\citet{Preibisch04} observed 30$\arcmin$ diameter Serpens area
with {\it XMM-Newton} for a total exposure time of 52 ks,
while \citet{Giardino07} surveyed smaller area (17$\arcmin$ diameter)
with {\it Chandra} in a deeper exposure of 90 ks.
The latter data was independently analyzed by \citet{Winston07},
who found a 95\% agreement between their and Giardino's et al. yields.
There are several limitations to these X-ray surveys.
One is the smaller FOV that only covers the Main Core area.
Next is the lower yield of detections for embedded objects.
For example \citet{Preibisch05} report $>$97\% detection rate
for the optically-visible T Tauri members of the Orion Nebula cluster,
while \citet{Winston07} only $\sim$50\% for classes 0/I-II in Serpens,
and no confirmed detections of class 0 is reported anywhere
(either due to high extinction, a flaring pattern of their X-rays,
or because X-rays are simply not emitted).
Sensitivity also limits the detection of the low-mass YSOs.
\citet{Winston07} find that the X-ray detection rate
in Serpens drops quickly below $\sim$0.2 M$_{\sun}$
for an age $\leq$3 Myr and extinction $A_{V}=$5.
The X-ray detected YSO sample can be still contaminated
by the foreground active binaries and M dwarfs. 
Finally, Figure \ref{flam_big} shows that there are
many X-ray sources whose nature can not be reliably identified
by means of the IR photometry. Spectroscopy may help clarify
whether they are faint members or contaminants.  

\citet{Oliveira09} followed with the optical spectroscopy
on the IR excess objects (from the c2d survey)
in the cluster B and the VV Ser region.
Out of 78 stars they identified 58 as cloud members
and 20 as background giants, which represents
the less embedded half of their original sample. 
They obtained a somewhat old median age
of 4.7 -- 7.5 Myr (depending on the models),
and a substantial age spread of $\gtrsim$15 Myr.
Considering their large errorbars, however,
the latter result should be taken with caution.
\citet{Winston09} classified 65 YSOs in the Main core area,
using optical and $HK$ NIR spectroscopy. Their sample
included both excess and non-excess members
drawn from the {\it Spitzer} and {\it Chandra} surveys,
as well as a four new members identified by the presence of the Li I line.
Seven objects were identified as probable contaminating giants.
\citet{Winston09} found a younger mean age $<$3 Myr for this upper region,
but also detected a tail of an apparently old population stretching to 20 Myr.
Younger objects were demonstrated to have a stronger clustering
and higher extinction than the older ones. This and the
detected decline of the SED slopes with age provided
confidence in the isochronal ages, despite large
uncertainties anticipated with the placement of YSOs on the HR diagram.   
These studies left open the question of whether
the older population was formed inside the current
pockets of the active star-formation and drifted out,
or whether there were different sites of star-formation in the past. 
\citet{Winston09} in addition found a surprisingly 
high disk fraction among the old cluster population.

In attempt to clarify these issues,
we carry an independent study of the Main core with FLAMINGOS.
In addition to obtaining new spectra and photometry,
we use spectral information in the de-reddening procedure
and in evaluation of the SEDs,
critically assess the distance to the cluster,
collate information on the individual sources
from the literature, and use our deeper FLAMINGOS images
to further examine the spatial distribution of the cloud population.

\section{Observations and Data Reduction}\label{red}

\subsection{Serpens Photometry}

Our Serpens observations were carried on two telescopes at the Kitt Peak
National Observatory, imaging on the 2.1 m telescope and spectroscopy on the 4 m one.
We imaged the Serpens molecular cloud in $JHK$ bands
with FLAMINGOS instrument in October 2003.
Seventeen $20\arcmin \times 20\arcmin$ fields
were tiled in the NE--SW direction to trace dense material
between $DEC(2000)=-0.5\degr .. +1.6\degr$, $RA(2000)=18^{h}32^{m} .. 18^{h}27^{m}$,
based on the CO and the 25 $\mu$m {\it IRAS} maps.
In this work we only discuss photometry for
our spectroscopic targets which are situated in the Main core
and were covered by a single field (Figure \ref{flam_big}).
The field was imaged on October 20 (in the $J,H$ filters) and on October 22 (in the $K$).
In the $J$ and $H$ filters 9 dithered exposures were taken 35 sec each,
in $K$ 25 exposures 15 sec each, for a total time of 5 -- 6 min per filter.
The plate scale was 0.6$\arcsec$ pixel$^{-1}$,
the seeing varied between 1.4$\arcsec$ -- 1.7$\arcsec$,
the airmass between 1.2 and 2.

The reduction was performed the same way as for the other
regions from our FLAMINGOS Molecular Cloud Survey \citep{Muench03, Levine06, Zuniga08}.
The raw images were reduced using {\sc IRAF}-based pipeline
developed by \citet{Zuniga06}. The PSF-photometry, astrometry,
and the source match between three filters were performed with
a pipeline written by \citet{Levine_thes}.
Finally, the zero-point of photometry in each filter
was corrected for geometrical distortion using an IDL-based routine
by A. Stolte. The image quality degrades away from the optical center,
which for this field is near $RA(2000)=18^{h}29^{m}40^{s}$, $DEC(2000)=+01\degr 09\arcmin$.
For each filter we plotted the difference between our and 2MASS photometry
as a function of distance to the optical center,
and corrected the trend with a 6th-order polynomial fit
(for more details see \citet{Zuniga08}).
Note that this procedure does not affect colors
that are defined by the filter system.
The $JH-HK$ diagram indicates that FLAMINGOS system is closer to the CIT than
to the 2MASS system.

The detection limits of our Serpens observations,
defined as the drop-off in the magnitude distribution function,
are around 18, 18, \& 17 mag
at $J, H$, \& $K$, respectively,
which is $\sim$2 mag deeper than 2MASS observations.
For spectroscopic sources FLAMINGOS errorbars
are 2 -- 5 times smaller than 2MASS,
particularly for the J band.
As a result, the presence of the NIR excess is now better
constrained on the $JH-HK$ diagram for faint sources,
while spectroscopically identified K--M background giants form
a tighter sequence.
FLAMINGOS photometry for our spectroscopic sample is reported in
Tables \ref{tableYSO} and  \ref{tableField}.

\subsection{Serpens Spectroscopy}

\subsubsection{Target Selection}

We selected our spectroscopic targets based on FLAMINGOS
$JHK$ photometry of the Main core region.
Objects with $K=10-15^{m}$ according to the evolutionary
models of \citet{Siess00} and \citet{bcah98} should have spectral types (SpT) of late-K -- M
for the estimated cluster age $\sim$2 Myr.
From our experience of observing embedded clusters
at similar distances (300 -- 400 pc) in Orion and Perseus
\citep{Luhman05, Levine06}, we should have been able to classify
M stars based on the depth of water bands at 1.4 and 1.8 $\mu$m
with a signal to noise ratio S/N$\ge$20 for a typical integration time of 1 hour.
We designed 4 masks with 30--40 slits each to cover the $10 \arcmin \times 10 \arcmin$
FOV of FLAMINGOS on the 4 m telescope,
centered approximately half way through
the dusty filament emanating south of the SRN (see Figure \ref{flam_big}).
The zoom-in on the area around the SRN is shown in Figure \ref{flam_zoom}.
Only 4 -- 6 objects per mask had a NIR excess
according to the $JH-HK$ diagram, the remainder
were picked uniformly across the field
according to the above mentioned magnitude cut.
However, due to the strong reddening in the region,
our $JH$ spectra of objects with $J>16.5$
turned out to be too noisy for a meaningful classification.
Similarly, the first 2--3 spectra from the top and bottom
(north and south) of the array
had to be discarded due to the poor signal in the flat fields.
In total, we obtained 30 good spectra, 27 of which
revealed spectral features strong enough
for spectral classification.
These objects are marked in Figure \ref{flam_big} and discussed
throughout the paper.

We further split our spectroscopic sample into two groups:
candidate YSOs and field stars.
YSOs are listed in Table \ref{tableYSO} together with
their identification numbers and the SED classes from the literature,
while field candidates are listed in Table \ref{tableField}.
Also given are the FLAMINGOS photometry and the c2d catalog numbers,
used for the construction of SEDs (\S \ref{sed}).
Throughout the paper we adopt for YSOs mainly the naming scheme
after \citet{Eiroa92} (EC), or after whichever study identified a particular source as
a YSO first, to be consistent with
the map of the cluster in Figure 2 of \citet{Kaas99}.
For the remaining stars (field candidates and source 19) we use
our own numbering sequence.
The YSO group includes objects that either have NIR or
mid-IR excess (SED classes 0 -- II),
non-excess objects (SED class III) that have been detected in the X-rays,
and source 19.
The field group consists of non-excess objects not known to be X-ray emitters,
and that either have been suspected
to be background giants from our spectroscopy,
or appear under-luminous/old on the HR diagram
(a possible foreground M dwarf with a small $A_{V} \sim 1^{m}$ star 28,
A-type stars 29 and 30, and a K-type star 27).

The SED class is usually assigned based on the slope
of the dereddened energy distribution $\alpha=\frac{\Delta log(\lambda F_{\lambda})}{\Delta log(\lambda)}$
between 2 -- 4 $\mu$m and 7 -- 14 $\mu$m.
The borderline values of $\alpha$ between class 0/I, flat (Fl), II (usually
corresponds to classical T Tau stars), and III are typically adopted as 0.3, $-$0.3, $-$1.6 respectively.
A finer subdivision is sometimes used, like transition disks (TD),
where excess only starts at wavelengths $\gtrsim$ 8 $\mu$m.
The general interpretation of the SED class is
that in class 0 objects the observed energy is originating entirely from
the optically-thick cold envelope, in class I objects a central source
becomes visible through a cavity in the thinned envelope,
in class II objects disk radiation dominates over the envelope,
and class III objects have largely dissipated
the primordial circumstellar material.
Since SED classification is not strictly defined,
as it relies on dereddening and sometimes on viewing geometry (like disk inclination,
cavity orientation, etc.), the object may be assigned
to two neighboring classes. We report SED classes in Tables \ref{tableYSO} and \ref{tableField}
as determined in the three main space-based
mid-IR studies of \citet{Kaas04}, \citet{Harvey06}, and \citet{Winston07}.
SED classes are shown in all of our figures to search for supporting evidence
of the described evolutionary interpretation. 

\subsubsection{Multi-Object Spectroscopy}

Spectra of the Serpens sources were obtained on the Kitt Peak 4m telescope
with FLAMINGOS with four masks on four nights
of 2004 October 4--7. Masks were designed based on the
FLAMINGOS $K$ band image obtained on the 2.1 m telescope and scaled
to the 4 m telescope FOV. The masks were $9.5\arcmin \times 3\arcmin$ 
in size, oriented north-south,
and were arranged along the right ascension near $RA(2000)=18^{h}30^{m}, DEC(2000)=+01\degr 10\arcmin$
to cover the area centered on the opaque
filament south of the SRN (Figure \ref{flam_big}).
Each mask contained 30 -- 40 slits, 0.95$\arcsec$ (3 pix) wide, 10--20$\arcsec$  long for the science targets,
and 3 -- 4 $8\arcsec \times 8\arcsec$ boxes for the bright alignment stars.
The mask design was optimized to produce as many non-overlapping
spectra as possible, one object per slit, and to allow
a 4$\arcsec$ nod along the slit for the sky subtraction.
The $JH$ grism coupled with the $JH$ filter provided
complete wavelength coverage at 0.9 -- 1.8 $\mu$m
with a dispersion 4.7 \AA/pix,
which resulted in spectral resolution of $R\sim$950 in the $J$ band and $R\sim$1250 in the $H$ band. 
Despite the diminished reddening in the $K$ band,
the $HK$ grism was not used for two reasons:
a strong, non-uniform background radiation
from the instrument
and the expected veiling in the $K$ band
for sources with NIR excess.
The latter effect occurs when the continuum radiation
from the circumstellar dust ($T_{eff} < 1500K$)
adds to the photospheric flux, diluting
weak spectral features, making it problematic
to measure line strengths in the low-resolution spectra.

Cluster observations consisted of three ABBA sequences,
where A and B correspond to the two positions of the star
on the slit separated by 4$\arcsec$, achieved
by nodding the telescope in the north-south direction.
Subtraction of the upper and lower exposures
removes telluric emission lines and the background radiation (from sky
and the instrument). A single exposure was 5 min long.
Due to occasional problems with read-out,
some AB pairs had to be discarded, resulting
in a total exposure time of 40 -- 60 min per mask.
Observations of the cluster field were followed by a few flat field 
quartz lamp exposures
through the mask, with the lamp on and off
(the latter is for the background subtraction).
An early G dwarf of $V\sim 6^{m}$ was then observed as a telluric standard
(HD 9562, HD198802, and HD 187923 for masks m1, m2 -- m3, and m4, respectively)
at an airmass within 0.2 of the cluster observation (except for m1,
where the cluster was at 0.5 larger airmass).
Standard spectra were obtained
in seven 5 -- 10 s exposures along a
3 pix-wide long slit, followed by
the flat-field exposures.
In some cases HeNeAr arc spectra were taken
to complement the wavelength calibration of the standard
from the OH telluric emission lines.
At the end of each night a series of dark frames were observed
with the corresponding integration times.
  
We reduced the spectra using Shrimpeater,
our {\sc iraf}-based pipeline.
Individual mask frames were dark-subtracted,
flat-fielded, and shifted for AB pair subtraction.
Two combined images were produced, the first of the slit spectra
with sky lines subtracted (in mask m3
some negative residuals can still
be seen in faint spectra),
and the second of the subtracted OH lines.
The spectrum of each source
(excluding alignment stars) was interactively traced and extracted,
the trace was applied to the sky spectrum from the same slit,
OH lines were identified, and the wavelength solution was applied
to the source spectrum.
The standard spectrum was reduced in a similar manner
(with the exception that the arc spectrum was used
in place of sky for wavelength calibration
of a mask m3 standard).
Telluric absorption lines in the standard spectrum
were matched against those in the target spectra
and a small shift (4--10 \AA) was applied to the
standard spectrum to adjust the wavelength calibration.
Target spectra were divided by the aligned standard spectrum
to remove telluric absorption (mostly water vapor),
visually examined to interpolate over noisy spikes,
and multiplied by the solar spectrum to restore the continuum shape.
Multiplication by the solar spectrum
is also done to compensate for weak intrinsic absorption lines in the standard.
To improve the S/N we further applied box-car smoothing
with the iraf task {\sc splot} and obtained the final resolution $R\sim500$.
Spectra of good S/N retained for the analysis
are shown in Figures \ref{fig_sp_under_yso} and \ref{fig_sp_under_fld}.

\subsubsection{Spectroscopic Standards}
 
For spectral classification of the program stars
we also observed with FLAMINGOS at Kitt peak 2.1 and 4m telescopes
a number of stars with known spectral types outside of the Serpens cloud.
These spectral standards are shown in Figures \ref{figMs} and \ref{figGKs}.
They were observed and reduced in a similar manner
as the Serpens targets. 

{\it IC348 standards.} IC 348 is a 2 -- 3 Myr cluster in the Perseus molecular
cloud ($d=$320 pc). Being less embedded, but otherwise
similar to the Serpens cluster, it was extensively studied in the optical
and IR and served as a benchmark for the investigation of properties
of young low-mass objects \citep[e.g.,][]{Muench07}.
We obtained FLAMINGOS spectra for a number of IC 348 stars
for which SpTs were determined from the optical spectra,
mostly by \citet{Luhman99} and \citet{Luhman03}.
We use the numbering and SpTs as adopted in Table 2 of \citet{Luhman03}.
F -- K stars and three early M stars (IC 348-65, 116, 92)
were observed in October 2004 in two masks.
The remaining M standards were observed in January and December 2003
in four masks and have already been
used for classification of other stars in IC 348 \citep{Luhman05}
and in NGC 2024 \citep{Levine06}.

{\it Field standards \& KPNO-Tau4.} To test the possibility that
some of the Serpens stars are unrelated interlopers,
we also obtained spectra of field stars with SpTs
available in the VizieR database\footnotemark.
\footnotetext{http://webviz.u-strasbg.fr/viz-bin/VizieR}
These stars and a late-type
young object KPNO-Tau4 were observed
through a long slit.
The A -- K standards are bright field stars that were
observed in January 2003 and October 2004 to serve as telluric standards
for other targets. Field M standards and KPNO-Tau4
were observed in October and December 2004
and presented in \citet{Levine06} and \citet{Levine_thes}.

Finally, for classification of the early-type stars,
we complemented FLAMINGOS standard star observations
with the $H$ band spectra of field stars from \citet{Meyer98}. 
The latter spectra were obtained with the Fourier transform spectrometer
(FTS) at the 4m Kitt Peak telescope and are available
through the NOAO archive\footnotemark. We convolved
them from the original spectral resolution of $R=$2900
to the FLAMINGOS resolution $R=$500.
\footnotetext{ftp://ftp.noao.edu/catalogs/medresIR/} 

\section{Spectral Classification}

We split our spectra into two groups:
those that show a steep break at 1.34 $\mu$m and a knee
at 1.68 $\mu$m due to water bands (13 sources), and those that do not (13 sources).
The former are candidates to be dwarf M stars and the latter
to be earlier types and giants. Because of the strong reddening,
the $J$ band is too noisy to reliably measure
spectral features, except for the 1.34 $\mu$m break.
Therefore, we classify the first group based mainly on the depth of the water bands
between the $JH$ and $HK$ bands (Figure \ref{figMs}), while the second group
based on the strength of a few lines in the continuum-flattened
$H$ band spectra (Figure \ref{figGKs}). Another 4 sources (EC117, EC129, 28, and H232)
are shown on both figures being borderline cases. 
We assign a range of spectral types to each
source based on the visual comparison with the spectral standards.
These SpTs are reported in our tables and in Figures \ref{figMs} and \ref{figGKs}.

\subsection{Classification of Candidate M dwarfs}

\citet{GL96} presented the first comprehensive
library of NIR spectra of YSOs in clusters.
Most of them must be late K--M stars according to the initial mass function.
This exploratory study revealed the challenges of identification
and characterization of YSOs. On the two-index diagram designating
equivalent widths (EW) of Na I $+$ Ca I and CO lines in the $K$-band spectra,
they found most $\rho$ Oph and R CrA YSOs situated in between the loci of
field dwarfs and giants, unlike in Taurus, where all YSOs except FU Ori stars
lay on the dwarf locus. No clear dependence on the SED class
was established. Further studies explored the role of veiling,
disk atmosphere, and age-related surface gravity ($log$g) effects
on the formation of the YSO spectra.
The conclusion was reached that veiling and
$log$g are the most important factors, and therefore for classification of YSOs
other YSOs should be used as templates, rather than field dwarfs.
Given that dust sublimation temperature is $\sim$1500 K,
veiling should have a smaller effect in our $JH$ spectra than in the $K$ band
often chosen by observers due to lower extinction. 
Due to low gravity in YSOs, matching to the field dwarf templates
can result in systematic errors in $T_{eff}$.
To create the first low-gravity templates, \citet{Luhman03}
constructed hybrid spectra by combining
observed spectra of a giant and a dwarf star of the same SpT.
They then classified a number of IC 348 members
based on these hybrid spectra to produce YSO templates.
We use the latter to classify our Serpens sources.
We also adopt the ``intermediate'' SpT-$T_{eff}$ scale of \citet{Luhman03}
for constructing the HR diagram (\S \ref{hr}),
since the scale was constructed specifically for a few Myr-old YSOs.

M stars with subsolar luminosities found in clusters with ages $<100$ Myr 
should be low-mass objects ($M<0.5M_{\sun}$) contracting
towards the Main Sequence (MS)
at nearly constant temperatures.
Large radii make their surface gravity
smaller than in old field dwarfs
of the same effective temperature ($log$g$=$3--4.5 vs. 4.5--5),
but not as small as in evolved red giants/supergiants
($log$g$=$0--3, e.g. Figure 6 in \citet{Ivanov04}).
While broad-band intrinsic colors of YSOs are dwarf-like
within the measurement errors, due to the gravity difference
their spectra past $\sim$M5
are clearly distinct from field dwarfs.
The $J$ band of field dwarfs is dominated by the strong absorption lines
of FeH and K I. In low-gravity objects these lines weaken and are replaced by
the shallower bands of VO and TiO
\citep{Gorlova03, McGovern04, Luhman07b}.
The difference is also seen in the shape of the $H$ band,
which is triangular in YSOs and more flat-topped
in field stars \citep{Lucas01, Allers07}.
Indeed, these differences are observed between the
IC 348 spectra and the field dwarfs in Figure \ref{figMs}.
Although in most Serpens sources we cannot confidently detect
lines in the $J$ band due to noise,
it appears that the $H$ band flux in all our late M candidates
rises from 1.50 to 1.68 $\mu$m, 
better matching the triangular shape of IC 348 sources
than the plateau of the field dwarfs.
We therefore regard all Serpens targets with
a strong water break to be young cluster members.
For early M stars it is more difficult
to claim membership based on our spectra alone.
Evolved stars (such as AGB, post-AGB, and red supergiants)
have extended molecular envelopes, often variable and chemically peculiar,
which results in a large diversity of their spectral shapes,
some of which resemble YSOs (e.g. \citet{Lancon2000}).
Being in addition luminous and numerous in the Galactic plane,
they can be confused with a YSO in the low-resolution spectroscopic studies
like ours. We rely on the X-ray detection to exclude this possibility
for our YSO candidates, except for the faintest and the most embedded ones.

We derived SpTs for Serpens sources by matching the depth of the 1.34 $\mu$m break
and the shape of the $H$ band to the IC 348 sources.
To eliminate difference caused by different reddening,
all spectra have been preliminary de-reddened
using {\sc iraf} task {\it deredden},
with a nominal value of $A_{V}$ that was selected
individually for each spectrum so that the slope of the line
connecting fluxes at $\sim$1.32 and 1.68 $\mu$m was $\sim$1.2 in all sources.  
In the early M stars the Mg I line at 1.50 $\mu$m
and the MgI/SiI doublet at 1.58 $\mu$m were additionally considered.
These lines weaken with later SpT until they disappear around M5.
We assign SpT M1 -- M8.5 to 17 Serpens objects
(excluding giant candidates, \S \ref{nowater}),
of which one star, 28, is likely a foreground dwarf
due to low reddening and the low position on the HR diagram (\S \ref{hrlabl}).

The errorbars that we report for our SpTs reflect the goodness
of the fit to the spectral standards.
Residual telluric features and veiling in the excess sources
can induce additional, systematic errors, that are also
more difficult to assess in the individual cases.
Based on our experience of observing similar object multiple times,
for a reasonable difference within 0.5 airmasses
between the targets and an F-G telluric standard,
the effect of the telluric residuals
in the water bands (within the wavelength intervals
shown on Fig. \ref{figMs}) should be no more than 1.5 subtypes.
Veiling can have a more significant effect on classification of excess sources,
in particular those with low S/N where we can not see atomic lines
(EC89, EC74, EC114, STGM1, H230, and [K99]40).
Our tests with veiling standard spectra showed that these YSOs
can be actually $\sim$2 subtypes later
if they are veiled by $r_{J}\sim0.5$.
Later types would make them even younger
than already observed. Accounting for the effect of veiling on the stellar luminosity will work
in the opposite direction, by making them fainter and older, but this effect is much smaller
than from SpT, as discussed in \S \ref{hr}.

There are 9 common sources with SpTs between us and
a higher-resolution study of \citet{Winston09}.
The latter study used two methods for classification --
visual comparison of broad features with standard spectra
and the ratio of the atomic line depths (that helps eliminate
the effect of veiling).
As shown in the Appendix, types agree between us and \citet{Winston09}
(within the stated errorbars) for 5 sources:
EC117($=$W216), W201, EC77($=$W204), [K99]40($=$W54), W68;
and disagree (by 2-3 subtypes) for 4: K407($=$W166), H232($=$W40), EC84($=$W85), EC86($=$W190).
There is no systematic pattern in the difference though,
from which we conclude that veiling is not significant
at least for these class II-III sources.

\subsection{Classification of Stars Without a Water Break}\label{nowater}

Three objects, all of SED class 0/I, lack sufficiently strong absorption lines
to allow a reliable classification:
SVS2, SVS20A, and EC 103.
These stars have steeply rising SEDs through the $H$ band without
a sign of overturn, which means either SpT $<$M1
or $A_{V}>20^{m}$. The $H$ band spectrum of EC103 exhibits hydrogen Bracket lines
in emission. The spectra of SVS2 and SVS20A
reveal extremely weak features that can be matched
with atomic lines in F -- K dwarfs. We did not attempt to examine these lines
in detail because at this level they may be affected by slight differences
between the telluric standard spectrum, the G0 -- G5 star HD 198802,
and the solar spectrum. The weakness of the lines indicates
a large degree of veiling, as is also suggested
by the location on the $JH-HK$ diagram red-ward
of the CTTS locus of \citet{Meyer97}.
We did not attempt to classify these stars,
but discuss them in the Appendix based on the information
available in the literature.

Stars 29 and 30 are the only clear cases
of hot stars in our sample having strong hydrogen lines in absorption.
Based on the strength of these lines and the absence 
of the He I line at 1.70 $\mu$m,
we estimate their SpTs to be between B5 and F2 \citep{Meyer98}.
The fact that the lines are broad indicates luminosity class V.
The location in the $JH-HK$ diagram
indicates $A_{V}=$ 6--8 mag for 29 and 8--10 mag for 30.
No excess emission is observed in the 24 $\mu$m
and 8 $\mu$m {\it Spitzer} detections of these stars, respectively.
They lie in the southern part of our spectroscopic field,
which is outside the area surveyed by X-ray telescopes.
Assuming SpTs B5, A0, and F2 for each of the stars and placing them
on the HR diagram as explained in \S \ref{hrlabl},
we observe that they lie $\Delta Log(L) =0.2-1.2$ dex lower than
the zero-age Main Sequence (ZAMS) of \citet{Siess00}.
We therefore regard these stars to be
field stars behind the cloud.

The remaining stars without prominent hydrogen lines
and water breaks are shown in the first column
of Figure \ref{figGKs}.
Other columns on that figure represent $H$-band
spectra of reference stars:
(1) FLAMINGOS spectra of YSOs in IC 348,
(2) FLAMINGOS spectra of field dwarfs and subgiants,
(3) FTS spectra of dwarfs and subgiants,
and (4) FTS spectra of low-gravity giants and supergiants. 
Examination of standards shows that
independent of the luminosity class,
SpT F is characterized by the presence of discernable hydrogen Bracket series,
that disappears by mid-G.
G type is characterized overall by a weak metal line spectrum,
which makes a classification difficult at low S/N. 
Mg I 1.50, 1.58, and Si I 1.59 $\mu$m lines dominate K type spectra,
reaching peak strength near K3. In dwarfs these lines disappear
by mid-M, in giants they are replaced by the CO second overtone band
and other molecules.

While it is easy to identify background M giants from M dwarfs by their strong absorption spectra,
in G -- K types it is more problematic due to a number of reasons:
in late-K types dwarf and giant temperature scales
coincide, in mid-G types $logg$ converge (at a solar value of 4.5 dex),
and on the $JH-HK$ diagram the loci of G -- K dwarfs and giants are inseparable.
This is a manifestation of the fact that most G -- K giants are lower mass or
less evolved stars compared to M giants, therefore their physical and spectral
properties are more similar to dwarfs. Another complication arises from
the fact that giants are observed over larger distances
sampling a wider range of metallicity,
and metal-poor giants have markedly weaker absorption spectra. 
Nevertheless, Figure \ref{figGKs} and examples in \citet{Ivanov04} indicate that
G -- K dwarfs can be recognized by the Mg I + Si I pair at 1.58 $\mu$m
with both lines of approximately equal strength (except that the Mg I doublet
becomes broader in early Ms), while in G -- K giants
Si I and the CO lines are more prominent.
The strength of the Mg I + Si I pair
in IC 348 sources resembles more that in dwarfs than giants.
A careful examination, however, reveals a prominent feature at 1.62 $\mu$m
that we tentatively interpret as a CO line, a signature of a reduced gravity in the YSOs.

We now consider the Serpens spectra. Objects 24 and 25 (marked ``g'')
have a clear match with late K -- early M giant/supergiant spectra.
X-ray non-detection and the excess-free SEDs facilitate
background giant interpretation for these stars.
We found two cases in the literature when giants
were confused with YSOs in Serpens based on the high extinction
and the geometrical association with the cloud --
CK2 (W158) in Figure \ref{flam_zoom}, and the neighboring object to
the protostar ESO H$_{\alpha}$279
in the north-western sub-core.
The giant nature was uncovered
through the observation of the strong CO absorption
in the $K$ band spectra of these two stars \citep{Aspin94, Chiar94, Casali96}. 
Objects 20, 21, and 22 also have strong features at 1.62 and 1.64 $\mu$m
resembling CO lines in the FTS spectra of mid-K giants.
On the other hand, they could potentially be a result of a very young age.
Objects 23 and 26 have a very prominent Si I line,
based on which they can be G -- K giants;
based on the weakness of Mg I lines, they could be G-type YSOs.
The spectrum of star 27 has the most dwarf-like appearance,
with only marginally enhanced CO lines.
We mark all these tentative giant candidates with ``g?'' (except 27),
but proceed treating them as YSOs for the placement on the HR diagram,
to see whether their age is compatible with the
age of the better established cluster members.
The flattened spectra of EC117 (an X-ray source)
and EC 129 (protostar according to the SED)
are consistent with the K -- early-M YSO interpretation.

Summarizing, among 13 red stars without the water break
we spectroscopically identified 2 A stars,
2 M giants, and 5 possible G -- K giants,
all likely being reddened background stars.

\section{HR Diagram}\label{hrlabl}

Our main goal is to derive the age of the Serpens population
and to compare it to other starforming regions.
This is achieved by comparing positions of sources
on the HR diagram
with model evolutionary tracks.

\subsection{$JH-HK$ Diagram and the Reddening Determination}\label{dered}

Because our sample is highly reddened,
optical photometry is not available for most of the sources,
while at \textit{Spitzer} wavelengths
one expects contributions from the circumstellar material.
The NIR is therefore an optimum wavelength regime to
study stellar properties of embedded objects.
We consider the location of our spectral
targets on the $J-H$ versus $H-K$ diagram
to clarify their membership, to identify sources with NIR excess,
and to estimate reddening for the subsequent determination
of the luminosity.

It is vital to understand where various groups of stars lie
in the $JH-HK$ diagram. First one needs to examine and tabulate the location
of non-reddened stars as a function of spectral type and luminosity class.
The \citet{Bessell88} compilation still remains the most popular
source of the NIR intrinsic colors of stars.
It was subsequently refined for main sequence stars
by \citet{KH95} in their
seminal study of the nearest star-forming Taurus-Auriga complex.
\citet{KH95} demonstrated that non-excess
low-mass YSOs are better dereddened to the main sequence
rather than the giant locus, despite the subgiant gravities
and an intermediate temperature scale \citep{Luhman99,Luhman03}.
\citet{Luhman99} arrived at the same conclusion for young brown
dwarfs (corresponding to spectral types $>$M5), based on the
comparison with colors of M stars belonging to the Galactic ``young disk''
from \citet{Leggett92}.
We adopt the locus of \citet{KH95} for B--K dwarfs,
and of \citet{Leggett92} and \citet{Luhman99} for M dwarfs,
in their original photometric CIT system,
as the photospheric colors for Serpens YSOs,
with the estimated accuracy of 0.05 mag.
Figure \ref{fig_jhhk} shows the NIR color-color diagram
for our spectroscopic targets.

One can see that objects suspected to be background giants
on the basis of our spectroscopy and the lack of X-rays
on average lie to the left from YSOs,
consistent with the interpretation of being G -- M giants
reddened by A$_{V}=6-17$ mag.
Giants/supergiants in general form a blue envelope (in $H-K$ color) on this diagram,
branching off the dwarf locus at $\sim$K5.
This is due to their extended atmospheres
favorable for molecule formation,
and in the case of the asymptotic branch stars
due to the modified chemical composition (e.g., carbon stars).
Giants can be seen to large distances and
are recognized on the Serpens diagram for the whole region
as a band of stars stretching along the upper-most reddening vector.
All spectroscopic giant candidates from our sample
lie in that band. However, three YSOs of SED classes II and III
fall in that region as well.
As a sanity check we place all giant candidates on the HR diagram
(except for the definite giant interlopers 24 and 25)
by dereddening them to the dwarf locus and assuming
them to be the early-type members of the cluster,
to see whether in this case their ages would be compatible
with the ages of the better established members.

On the other hand, class 0/I and flat objects
(except for EC129) lie to the right
of the reddened stellar locus, together with
the majority of class II.
Their reddening, however, is on average
larger than in class II and III objects.
Only two sources, class II EC84
and class III EC86, have a location
inconsistent with their late M SpT,
having either a $J-H$ color too red
or $H-K$ too blue. We discuss them further
in \S \ref{sed} and the Appendix.

The $JH-HK$ diagram confirms that the
reddening is non-negligible in the region
and needs to be accounted for in the photospheric luminosity estimation.
It can also affect the SED classification.
Without knowledge of SpT, however, it is non-trivial
to separate reddening from the excess NIR emission.
\citet{Kaas04} and \citet{Djupvik06} determined reddening
by adopting for all sources
one value for the intrinsic color $(J-H)_{0}=0.85$, an average
value on the CTTS locus. \citet{Harvey07b}
on the other hand neglected excess contribution in the NIR wavelengths
entirely by adopting for $(J-K)_{0}$
a photospheric color of a K7 dwarf star (an average SpT of CTTSs).
We determine reddening by examining both location
on the $JH-HK$ diagram
and the SpT.

We use two methods for dereddening spectral sources,
depending on whether they deredden to the
dwarf locus approximately at their SpT or redward of it,
indicating presence of the NIR excess.

For all class III sources (except for giants 24, 25
that we did not attempt to classify)
and three class II (EC84, W68, K407) we use $J-H$ color rather than $H-K$.
Even if a small NIR excess is present in the latter sources,
because the dust temperature is $<$1500 K
the excess will be less in the $JH$ bands
than in the $K$ and will not affect our $A_{V}$ estimate significantly.
This fact also explains our choice
of $J$ magnitude for deriving photospheric luminosity (\S \ref{hr}).
Using the reddening law of \citet{Cohen81}, we adopt

{\footnotesize
\begin{equation}\label{eqn_av1}
A_{V}=\frac{A_{V}}{E(J-H)} (J-H-(J-H)_{0})=9.09(J-H-(J-H)_{0})
\end{equation}
}

\noindent where $J-H$ is the observed color and $(J-H)_{0}$
is the expected intrinsic color, based on the SpT of the object
and the SpT--$(J-H)_{0}$ dwarf calibration of \citet{KH95} for B--K stars
and of \citet{Luhman99} for M stars.

For sources with strong NIR excess and
the less obvious NIR cases with excesses
at {\it Spitzer} bands (EC89 through [K99]40 in Tables
\ref{tableYSO} and \ref{tableProp}) we determine
$A_{V}$ from the $J-H$ and $H-K$ colors by
dereddening them along the reddening vector to the {\it adjusted} CTTS locus. 
The CTTS locus shown on Fig. \ref{fig_jhhk}
was defined by \citet{Meyer97} for low-mass YSOs in Taurus
with typical SpTs K7--M0.
\citet{Luhman99} and \citet{Liu03} among others, however, showed that
many cooler stellar/substellar young objects lie
below this classical TTau star locus, and suggested
that its origin on the dwarf locus
depends on the SpT of the source.
By calculating the intercepting point of the reddening vector
passing through a given source
with the CTTS locus that was shifted to originate on the dwarf locus
at the SpT of that source,
we arrive at the following
expression for $A_{V}$ for stars with the NIR excess:

{\footnotesize
\begin{eqnarray}
\nonumber
\nonumber A_{V}=\frac{\frac{A_{V}}{E(J-H)}}{\frac{1}{k_{CTTS}}-\frac{E(H-K)}{E(J-H)}} \left(\frac{J-H-(J-H)_{0}}{k_{CTTS}}
-(H-K-(H-K)_0)\right)
\end{eqnarray}
\begin{eqnarray}
=13.831(J-H-(J-H)_{0})-8.022(H-K-(H-K)_{0})
\end{eqnarray}
}

\noindent where $k_{CTTS}=0.58$ is the tangent of the slope of Meyer's CTTS locus in the CIT system,
and all other designations are as in (\ref{eqn_av1}).

Table \ref{tableProp} lists the extinction and other photospheric parameters
that we derive based on our SpTs for YSOs (giant candidates were omitted).
Errors in $A_{V}$ were obtained by quadratically propagating
uncertainties in the observed and intrinsic colors
in expressions (1) and (2).
The uncertainty in the intrinsic color results from
a range of SpTs derived for a given source
and a 0.05 mag uncertainty in the standard color for a given SpT.

\subsection{Distance}\label{dist}

Accurate Hipparcos parallaxes are not available for
this embedded area, requiring indirect techniques for distance determination.
Two distance scales to the Serpens cloud are used in the
modern literature.
In their (sub-)millimeter study \citet{Hoger99}
adopted 400 pc based on the 425 $\pm$45 pc estimate
of \citet{Chiar97}. The more recent radio and {\it Spitzer} surveys,
however, use a shorter scale of 220 -- 260 pc,
based on two studies of \citet{Straizys96} (259 $\pm$ 37 pc)
and \citet{Straizys03} (225 $\pm$ 55 pc).
An intermediate value of 311 ($\pm$ 38) pc was adopted in the binary surveys
of \citet{Haisch02} and \citet{Haisch04},
based on the earlier estimate of \citet{deLara91}.
What is the reason for the inconsistency among these estimates?

The estimate of \citet{deLara91} is based
on the spectrophotometric parallax of 5 B -- A stars
from \citet{Chavarria88} that have been associated with
the cloud because they are surrounded by reflection nebulosities.
These stars are: 
R1 (2MASS J18213334-0202295), R7 (2MASS J18295606+0100228, next to IRAS 18273 $+$0059), 
R12 (HD 170739), R13 (HD 170784), and R15 (HD 171491).
Note that only R7, R12, and R13 lie within the boundaries of the Main cloud,
as can be seen in the IRAS 60 $\mu$m map of \citet{Zhang88a}.
The other two stars are situated further south, outside the area
surveyed in our work and with {\it Spitzer}.
The distance was obtained by \citet{deLara91} in the following manner.
From the reddening-independent $[u-b]$
and the gravity-sensitive $[m1]$ photometric indices,
complemented by spectral types
from the medium-resolution spectra,
they determined effective temperatures
and the dwarf luminosity class (class V) for these stars.
The reddening was obtained by comparing
the observed optical--NIR colors
to the model atmospheres of \citet{Kurucz79}.
Visual magnitudes were corrected for reddening,
and the distance was obtained by comparing them to
the standard absolute magnitudes ($M_{V}$) as tabulated
in \citet{SchmidtKaler82} as a function of SpT.
\citet{Straizys96} found that two of de Lara's et al. (1991) stars,
R12 and R15, are nearly equal flux binaries,
which means their distances were underestimated in \citet{deLara91} by $\sim\sqrt{2}=1.4$ times.
Averaging de Lara et al.'s (1991) distances for
the remaining stars R1 , R7, \& R13
(337, 416, and 330 pc respectively), we obtain 361 $\pm$ 34 pc.

\citet{Chiar97} determined distance using 7 of the \citet{Chavarria88} stars:
R1, R7, R10 (HD 170634, illuminates S68), R12, R13, R15, \& R16 (2MASS J18365749+0020357).
Similar to de Lara's et al. (1991) study, she employed
spectrophotometric parallaxes.
R10 was reclassified from A1 to B7 based on UBV photometry.
For the remaining stars she adopted the
SpTs of \citet{Chavarria88}/\citet{deLara91}.
$M_{V}$ was determined as an average from two methods:
the $\beta$-index -- $M_{V}$ calibration of \citet{Crawford78} (using photometry of \citet{Chavarria88})
and the SpT -- $M_{V}$ calibration of \citet{SchmidtKaler82}.
She then examined the location of these stars on the $JH-HK$ diagram
to conclude that none possesses circumstellar material
that could modify the interstellar medium (ISM)
value of the total to selective extinction
ratio $R=\frac{A_{V}}{E_{B-V}}$, which was found to be 3.1 $\pm$ 0.1
in agreement with 3.3 $\pm$ 0.3 of \citet{deLara91}.
If we omit: the shell star R10 that can be more luminous than MS stars,
for example, \citet{Houk99} assigned luminosity class IV for it;
the binaries R12 \& R15;
R16, discarded by \citet{deLara91} due to
the outlyingly-large distance $>$600 pc with a small $A_{V}=2.7$,
then Chiar's (1997) distances for the remaining sources
are 390 pc, 377 pc, and 553 pc for R1, R7, and R13, respectively.
We note that R13 is of the earliest SpT of the three, being a B3 star.
An uncertainty in $\beta$ index $\sim$0.02 mag, corresponding
to 1 spectral subclass, at B3 translates to uncertainty of about 0.5 mag in $M_{V}$,
or 20\% in distance,
but only of 0.2 mag and 10\%, respectively, at A2 (star R1).
Omitting R13 and averaging distances for R1 and R7, we obtain 383 $\pm$ 9 pc.

The sample of \citet{Straizys96} is the largest: 18 stars lying
within $1\degr$ of S68 (R10, or star 61 in their notation).
Spectral types, $M_{V}$, and reddening were obtained
in that work simultaneously
by fitting seven-color optical photometry
to the calibrated grid of intrinsic colors
for $\sim$100 stars from the region down to $V=13^{m}$.
A plot of $A_{V}$ versus distance was then constructed
by plotting individual values for each star.
The near side of the cloud was determined at
the abrupt upturn of the extinction values at $d\sim$200 pc.
The far side was defined where no star of zero extinction was observed, at $\sim$300 pc.
Distances to 18 stars with $A_{V}>1^{m}$ within this distance range
have been averaged to obtain 259 $\pm$ 37 pc.

There are five stars in common between \citet{Straizys96},
\citet{deLara91}, \& \citet{Chiar97}:
R10, R12, R13, R15, and VV Ser.
Stars R10 and VV Ser should not have been used by \citet{Straizys96},
since they are an evolved shell star
and a pre-Main Sequence (PMS) Herbig AeBe star, respectively.
If we compare distances to the remaining stars (after multiplying the distances
of \citet{deLara91} and \citet{Chiar97} by 1.4 times for the binaries R12 and R15),
we find that the distances
of \citet{Straizys96} are systematically lower.
The reason for the disagreement for these stars lies
not so much in the reddening but rather in the absolute magnitudes:
the values of $M_{V}$ of \citet{Straizys96} are systematically fainter.
\citet{deLara91} and \citet{Chiar97} used SpT/color calibrations
for dwarf (luminosity class V) stars, while \citet{Straizys96} for
ZAMS stars.
We note that both calibrations, of \citet{Straizys96} (as published in \citet{Vil82})
and of \citet{SchmidtKaler82}, agree well within a given luminosity type.
Since massive stars spend considerably less time on the ZAMS,
for a random hot star in the field the use of a dwarf calibration
is more appropriate. ZAMS stars are fainter than the more
evolved dwarfs, leading to
the underestimation of the distance in \citet{Straizys96}.
At B6 (star R7) the difference is 0.8 mag, 
leading to a 40\% difference in distance, at A2 (R1) it is
0.5 mag (20\% in $d$), and only by F0
$M_{V}$-s become comparable.

In the subsequent paper that explored other sub-areas of the Aquila Rift
\citet{Straizys03} indeed adopted a dwarf calibration
for luminosity class V (judging by the catalog given in \citet{Balta2002}).
A caution, however, should be exercised when applying
results of their shallow optical surveys
to the embedded regions like the one studied here.
For example, $A_{V}$ does not exceed 3$^{m}$ on their extinction plots,
while already the earliest NIR studies of the Serpens cluster indicated
typical $A_{V}\sim 10^{m}$.
At high extinction and large distances
only bright A--B stars are accessible;
their luminosities, however,
are very sensitive to the uncertainties in age and SpT,
as discussed above.
Furthermore, \citet{Straizys03} regard 225 ($\pm$ 55) pc
as only the front edge of the cloud complex,
while the location of the back side is far less certain
("thickness can be about 80 pc").

Finally, \citet{Oliveira09} cite a private communication
with Jens Knude (2008) that claims distance of only 193 $\pm$ 13 pc.
With this distance a median age for their IR excess sources
would be 10--16 Myr (depending on the tracks),
which is quite old considering a well established
primordial disk dissipation time-scale of $<10$ Myr.

For this paper we adopt a distance
to the SRN cluster of 380 ($\pm$ 50) pc,
which places it toward the back of the Aquila complex.
Our investigation demonstrates that the distance to the Serpens
star forming region is still not well constrained and
requires a dedicated study.

\subsection{Temperatures and Luminosities}\label{hr}

Luminosities of spectrally classified Serpens sources
were derived from the dereddened $J$ band magnitude.
Using the reddening law of \citet{Cohen81}:

{\footnotesize
\begin{equation}
log(L/L_{\sun}) = 1.86 - 0.4(J-0.265A_{V}+BC_{J}-5log(380)+5)
\end{equation}
}

Bolometric corrections $BC_{J}$ are determined for each SpT
from the $BC_{V}$ and $V-J$ dwarf compilations of \citet{KH95} and
\citet{Luhman99} (K. Luhman 2001, private communication).
The uncertainty in $BC_{J}$ for a given SpT
is estimated to be 0.1 mag, from comparison between the values
of \citet{Luhman99} and \citet{WGM99} for M dwarfs.
The uncertainties in luminosity were estimated similarly
to the uncertainties in $A_{V}$, by quadratically adding
the uncertainties of the two groups of variables that
depend on the observed photometry and on SpT. 

Temperatures for F--K stars were obtained from the SpT--$T_{eff}$
dwarf relation of \citet{KH95}, while for M stars the ``intermediate''
relation for YSOs of \citet{Luhman03} was used. The latter
is constructed to provide an intermediate value of $T_{eff}$ for a given
SpT between the hotter giant and the cooler dwarf values.
The largest difference between the giant and dwarf
scales of $+$500 K is reached at M6.
At M0 the difference disappears
and in K types reverses its sign.
The scales were combined by \citet{Luhman03} to fulfill
the condition of coevality of the members of the young quadrupole system GG Tau
when using PMS evolutionary tracks of \citet{bcah98} (BCAH98).
The error of $T_{eff}$ for each source in Table \ref{tableProp} only
reflects the uncertainty caused by the error in SpT.
As long as one considers the age spread in the cluster
or compares mean ages between clusters
studied with the same temperature scale
and evolutionary tracks, the systematic errors in the scale
should have a negligible effect. 

Figure \ref{hrBCAH} shows Serpens sources over-plotted on
the frequently used BCAH98 tracks.
These tracks are based on the dust-free NextGen atmospheric models \citep{Hauschildt99a};
they only start to deviate from CBAH00 DUSTY models \citep{Chabrier00}
at ages older than 10 Myr and cooler than about 2300 K,
which is outside the range of our targets.
For consistency, we plot BCAH98 tracks above 0.6 $M_{\sun}$
with the same mixing length parameter as for lower mass stars,
$\alpha_{mix}=1.0$.
We note that this value of $\alpha_{mix}$ may overestimate masses
of EC117, EC 129, and some giant candidates
by up to 30\%, as compared with the recommended $\alpha_{mix}=1.9$.
This difference is, however, within the large
error-bars on SpTs of the early type members.

The earliest available isochrone in BCAH98 models
is 1 Myr, and a large fraction of our sample
falls above it. Models of \citet{DM97} (DM) provide
isochrones as young as 0.1 Myr, though they may be less accurate
being based on the gray atmosphere approximation.
In Figure \ref{hrDM} we show the latest 1998 update of these models.
Three sets of tracks are available with different initial deuterium abundance fractions.
We chose the median value of $2\times10^{-5}$ as it reaches to larger masses;
the differences are insignificant in our mass range anyway.
The tracks of \citet{DM97} originate from the so-called birthline,
that in this case is defined at the on-set of the deuterium burning.
All our YSOs except for EC84 fall at or below this line within the errorbars,
lending support to their membership in the cloud and accommodating
our larger distance. Equal-mass binarity (see Appendix),
or an earlier SpT could bring EC84 down to the birthline as well.

Masses according to the two sets of tracks are reported in
Table \ref{tableProp}.
Above 1 Myr isochrone (0.1 Myr in DM97) each track was
extended based on the approximately linear 
interpolation of the the track's segment between 1 and 3 Myr.
The errors in the mass and age for these objects were adopted similarly to the
objects that lie within the model grid, by finding the intersection of
the errorbars with the the nearest interpolated isochrone/track.
Evolutionary tracks of low-mass YSOs
are nearly vertical, therefore, the error in the BCAH98 mass
in Table \ref{tableProp} is determined primarily by the error in SpT.
DM98 masses are given for comparison only.
DM98 and BCAH98 tracks best agree between 0.1--0.25 $M_{\sun}$
and 2--10 Myr, at the location of W201.
At masses $M>0.2 M_{\sun}$ DM98 tracks predict 2--3 times smaller masses
and younger ages than BCAH98 tracks, while at $M<0.1 M_{\sun}$
they predict slightly larger masses and ages.

Evolutionary masses allow us to estimate surface gravities
that will be used in selecting photospheric models
for fitting SEDs (\S \ref{sed}):

{\footnotesize
\begin{equation}
log(g)=4.42+log(M/M_{\sun})-log(L/L_{\sun})+4log(T/5770)
\end{equation}
}

It is difficult to analytically derive errors on l$og$g from this expression,
considering the non-linear dependences of mass and luminosity
on magnitudes and SpT. The difference between $log$g obtained
with BCAH98 and DM98 masses, and the difference between $log$g obtained
for an upper and lower limit on the SpT for a given set of tracks
are comparable to each other and are typically 0.1--0.3 dex.
The largest uncertainties are for EC129 and EC117, due to differences
in the tracks at the high mass end.
These uncertainties are still within the 0.5 dex step
of the NextGen synthetic spectra that we use to fit the SEDs.

Since M2-M3 types of \citet{Winston09} for EC84 and EC86
better explain their NIR colors, we re-calculated their positions
on the HR diagram (cyan squares on Figs \ref{hrBCAH}, \ref{hrDM}).
Both objects remain younger than 1 Myr, but EC84 now moves under the birthline,
which makes it a more likely member.
This exercise illustrates that unaccounted uncertainties in the SpT M of 3 subtypes
translate to the uncertainty in age of the order of the distance between the isochrones
drawn on our HR diagrams. The effect is largely
due to the change in the temperature, while luminosity is barely affected.

\citet{Winston09} discuss systematic uncertainties on luminosity (for class II objects).
The strongest effects are due to the scattered light
and accretion in strongly embedded or inclined sources, resulting in up to 1 dex scatter in the deduced $logL$.
We can estimate the effect on the stellar luminosity
of the unaccounted excess emission
from the hot dust in the disk.
It will overestimate photospheric luminosity directly,
as well as through the overestimation of extinction (as both excess and extinction make object look redder).
To estimate this effect on our luminosities, we can re-write the expression
in \citet{Winston09} for the $J$ band:
{\footnotesize
\begin{eqnarray}
\Delta logL(exc)= -0.4 \times ( \Delta m_{J}(exc)-[J-H]_{exc} \times A_{J}/E(J-H) )
\nonumber
\end{eqnarray}
}
Adopting $<m_{J}(exc)>=-0.26$, $<m_{H}(exc)>=-0.47$ from \citet{Winston09}/\citet{Cieza05}
(corresponding to $r_{J}=0.27$, $r_{H}=0.54$), and $A_{J}/E(J-H)=2.41$ from \citet{Cohen81},
we obtain 0.10 dex for the first term in the above expression and 0.20 for the second.
Similar effect is obtained when using $H$ magnitude for luminosity determination instead of $J$ \citep{Winston09}.
However, since in our work de-reddening was performed
to the excess locus, the last term can be omitted from this calculation,
resulting in the increase of luminosity due to veiling in the $J$ band of only 0.1 dex.
Uncertainty in SpT, therefore, remains the dominant factor
that affects the position of class III and
the majority of class II objects on the HR diagram.

\section{Spectral Energy Distributions}\label{sed}

Unresolved warm circumstellar dust around a YSO
can be studied through examining the
spectrum and amount of emission it produces in the IR.
This excess emission has first to be corrected for
the interstellar extinction and
separated from the photospheric flux.
Our individually determined photospheric parameters allow us
to carry out this procedure more accurately
(in particular in the cases of small excesses
such as evolved disks) and to search for correlations
with stellar properties.

We retrieved {\it Spitzer} measurements from the High Reliability
Serpens catalog (5th data release) of the Cores to Disks Legacy team,
and merged them with FLAMINGOS photometry to construct SEDs.
The raw and dereddened (following \citet{Mathis90}) fluxes are plotted
in Figure \ref{figsed}. The stellar contribution
in the SEDs is represented by the smoothed solar-metalicity
NextGen model spectra of the PHOENIX group \citep{Hauschildt99a} .
The physical parameters of the model spectrum for each star ($T_{eff}, logg$)
are the nearest in the grid to our spectroscopically determined values
from Table \ref{tableProp}.
The $A_{V}$ values used for de-reddening of the observed SEDs
are our values from Table \ref{tableProp} (see \S \ref{dered}).
We emphasize that these parameters are based on SpTs and assume YSO properties:
an intermediate temperature scale and the dwarf intrinsic colors.
NextGen models ($logg \geq 3.5$) are drawn
in panels a), b) (except for giant 24), and the first panel of c).
For suspected field giants in addition
an alternative NextGen-{\it giant} \citep{Hauschildt99b} set of models
is shown on the second sub-panel of panel c). The giant models involve
spherical geometry to treat the extended atmospheres.
To select appropriate giant models we adopted
the giant SpT--$T_{eff}$ scale of  \citet{Perrin98},
and dereddened those SEDs with $A_{V}$ derived from the giant SpT--$(J-H)_0$
calibration of \citet{Bessell88}.

If the dereddened {\it Spitzer} points lie above the model photospheric flux,
it can indicate either an IR excess or incorrectly
determined photospheric parameters. If they lie below,
it can only be due to the latter reason,
unless the models produce systematically wrong colors.
The small over-prediction of flux in the $H$ \& $K$ bands at $<$10\% level
(0.05 dex in log $\lambda$F$_{\lambda}$) seen for sources with $T_{eff}<$4000 K
is a known feature of NextGen models, and is likely caused by the incomplete treatment
of water bands \citep{Leggett96, Leggett01}.
Except for this effect, no systematic offsets
are seen between the models and {\it Spitzer} fluxes in Figure \ref{figsed}.
Among YSO candidates we find only two cases where
SpTs and the assumption of dwarf NIR colors
result in photospheric parameters inconsistent with their SEDs: EC84 and EC86. 
As discussed in the Appendix, the discrepancy is likely due to earlier SpTs.
Models corresponding to M2-M3 types of \citet{Winston09}
can fully account for the dereddened fluxes
of these objects up to 6 and 8 $\mu$m in EC84 and EC86, respectively.
 
The last panel of Figure \ref{figsed} verifies the giant hypothesis for
the spectroscopically suspected giants.
As one can see, the SEDs of all ``g?'' sources
can be described equally well when applying YSO
and giant K3 -- K5
SpT--$T_{eff}$--$(J-H)_{0}$ calibrations.
For the strong M giant candidates 24 and 25, however,
the dwarf $(J-H)_0$ can be firmly ruled out,
requiring $A_{V}$ that is too large to fit the SEDs.
These two objects are better fitted with M0 -- M3 giant models
(only the giant fit is shown for source 24 in Figure \ref{figsed}).
Sources 23 and 26 could be G type stars according to their spectra.
For source 23 a good match is obtained with both dwarf and giant models
of SpT G7, while for 26 only a fit with a K type model is possible
(G type requires too large $A_{V}$ inconsistent with the SED).
We conclude that the SEDs of all G -- K field candidates
are consistent with the background giant interpretation.

For stars showing excess above the photosphere we compare
their de-reddened SEDs with the SpT-dependent
median thick- and thin(``anemic'')-disk SEDs of IC348 YSOs from Table 4 of \citet{Lada06},
de-reddened by $A_{V}=2.5$ as recommended by the authors.
Dependence of SED on the SpT has been largely ignored in the past studies,
by using Meyer's et al. (1997) approximation of the CTTS locus 
on the NIR color-color diagrams or
the median CTTS SED from \citet{Dalessio99}.
Both have been based on the study of the solar-mass YSOs
from the Auriga-Taurus star-forming region,
with typical SpTs of late-K and ages 1--2 Myr.
{\it Spitzer} surveys of various star-forming regions
provided enough statistics to refine SED classification.
According to the IC348 templates,
the largest NIR excess is observed in K stars, followed by
G and M stars, and finally by A--F stars.
Note, however, that the SEDs of Lada et al. are empirical,
so it is not clear whether the reflected
dependence on SpT is due to different illumination
by the host star \citep{Muzerolle06} or due to the different
evolutionary stages, e.g. dust settling.
By examining the fraction of stars with excesses
and gas accretion indicators
as a function of cluster age, it has been shown
that the low level of the NIR excess in massive stars
must be due to the shorter disk dissipation time
\citep{Haisch01a, Hernandez05, Carpenter06, Lada06}.
On the other hand, the low-mass stars hold to their 
primordial disks longer,
but the NIR excess is suppressed for the following reasons:
1) the peak of the photospheric emission shifts
to longer wavelengths decreasing
the excess contrast, 2) the disks themselves are
expected to be cooler and smaller.
Figure \ref{figsed} shows a good agreement between
D'Alessio et al. SED (in red) and a thick disk
SED of Lada et al. (green) for K6--M0 stars ($T_{eff}=$4200--3800 K),
as expected. The majority of our YSOs, however,
have later SpTs and should be compared to the Lada et al's SEDs
(see a case-by-case discussion in the Appendix).

\section{Discussion}

\subsection{Circumstellar Properties of Serpens YSOs}

As shown in the Appendix, the physical properties
and the accurate evolutionary stage
even for the brightest IR sources in the cloud are
poorly constrained in the literature.
Using our photometry and spectroscopy, we derived
uniform SpTs, extinction, and photospheric luminosities for 16 YSOs
from the Serpens core cluster
and constructed the HR diagram.
For 5 YSOs (EC89, EC74, EC114, STGM1, H230, SED classes 0/I--II) and a YSO candidate 19 (class III)
SpTs were determined for the first time,
as well as for 11 field candidates.
Our sample of YSOs according to BCAH98 tracks turned out to be comprised mostly of
low-mass stars, 3 substellar objects,
2 solar-mass stars, and 2 intermediate mass stars (SVS2 and SVS20),
of various SED classes.
Eleven excess-free stars examined as possible Class III members
have been identified as probable interlopers, mostly background red giants.

There is a longstanding debate on the nature of the flat SED objects.
Their SEDs are intermediate between the rising class 0/I 
where excess is produced by the massive cold envelopes, and the declining class II
where it is produced by the un-obscured hotter disks. Does it mean that flat sources
represent an intermediate evolutionary stage?
Not necessarily according to \citet{Crapsi08}, who claim that
flat objects are disk systems seen edge-on.
Furthermore, \citet{White04} investigated high-resolution optical spectra
of class I and class II sources with HH outflows
and found little evidence of a systematic difference with the
more typical class II sources,
in terms of mass, luminosity, rotation rate, accretion rate,
and the millimeter flux density. The larger frequency of
the outflow-related forbidden emission lines
in class I objects in that study was explained by a better contrast against the
obscured photospheric spectrum in the edge-on disk systems.
On the other hand, \citet{Doppmann05} found
based on their high-resolution $K$-band spectra
larger $vsini$ and veiling in classes I-flat than in class II objects.
They argued that the sample of \citet{White04}
was not representative enough: it only covered the Taurus-Auriga region
and was biased toward
optically-visible, more evolved objects.
\citet{Beck07} analyzed $KL$ spectra of 10 variable class I and flat objects in Taurus.
Based on a number of diagnostics they concluded that
of the 4 flat sources ($\alpha$ between 0.10 and 0.25)
1 was a truly transition source, 1 a protostar, and 2 reddened CTTS.

What about Serpens flat SED objects?
First, we observe that they
can be found anywhere on the $JH-HK$ diagram
where $A_{V}\gtrsim6$, mixed with CTTSs and protostars.
IRAC-based classification helps single them out,
but these wavelengths are
still susceptible to extinction.
As a result (as shown in the Appendix),
studies that perform classification on de-reddened SEDs
tend to assign later classes than those that use
the observed SEDs.
It is difficult to estimate reddening
without the knowledge of the SpT;
not surprisingly, for some of our objects the $A_{V}$
values in the literature vary by as much as 5 -- 10 mag.
The key to typing YSOs in our work
was comparison of the absorption
features in their $JH$ spectra (whose origin we attribute
to the photosphere) with features in the the optically-classified YSOs
from a similar-age cluster IC348.
The protostars that we were able to classify
are all located within the reddened CTTS locus on the $JH-HK$ diagram,
indicating that excess emission could primarily originate from a disk.
Detection in the X-rays of all observed flat-class objects but EC129
also makes them more similar to CTTS than class 0/I objects.
Furthermore, after correcting for extinction
we found that all of the flat class objects
turned out to have SEDs resembling the CTTS SED of \citet{Dalessio99}
(except maybe for H232 with lower excess in the IRAC bands).
And while we could not deredden
SVS2 and SVS20, their mid-IR spectra (obtained thanks to their high luminosity)
also reveal signatures of disks (the silicate emission).
However, all our spectrally classified flat objects
(except EC129) turned out to be low-mass M stars,
which is not surprising considering that our low-resolution spectroscopy
is most favorable at characterizing late-type YSOs.
Compared to the CTTS disk of \citet{Dalessio99} that represents
more luminous solar-mass stars,
disks around M stars can be expected to show smaller excesses.
We therefore compared Serpens flat class SEDs with
the SEDs of low-mass stars from \citet{Lada06}.
The comparison revealed (Fig. \ref{figsed}) that Serpens
flat objects do have more excess at mid-IR wavelengths than 
the similar SpT class II objects in IC348,
perhaps indeed indicating an earlier evolutionary stage.
Serpens class II objects, on the other hand,
have comparable SEDs to their thick disk counterparts in IC348.

In two class flat/II objects (H232 and STGM1)
we detected Pa$\beta$ in emission, which is interpreted
as a sign of a gas accretion.
The accretion luminosities inferred for these stars (see calculation in the Appendix)
are average for CTTSs, as well as for class I \citep{Muzerolle98, Beck07}.
H232 is of particular interest as its low-level
NIR excess contrasts with the optically-thick mid-IR excess,
indicating a sort of clearing in the inner disk,
perhaps due to formation of a gas giant planet.
In later type members the accretion is not excluded,
but we could not detect comparable strength Pa${\beta}$ in their
lower S/N spectra.

We confirm the lack of excess in all class III {\it Spitzer}-classified objects.
No NIR or IRAC excesses are observed beyond 10 Myr,
in agreement with other studies
that found primordial disk fraction
of only few percent beyond this age \citep{Haisch01b, Hernandez08}.
The optically-thin emission at 24 $\mu$m, in particularly expected
from debris disks, can not be probed in our low-mass sample
being below {\it Spitzer} detection limit.
Our confirmation of the presence of the transition/evolved disks ([K99]40 and K407),
however, encourages searches of young thin disks among IRAC-classified class III
objects with the next generation of the IR facilities, such as {\it Herschel}.

In our study the previously reported SED classes were generally confirmed.
It is not always because $A_{V}$ simply turned out to be too small
($<$ 10 mag) to modify their SEDs.
We find that the interpretation of the SED classes
crucially depends on the knowledge of the SpT of the host star.
The amount and the shape of the deduced excess depends on how precisely
the reddened photospheric flux was subtracted, for which one needs to know
the temperature and (to a lesser degree) gravity of the star.
Furthermore, for a given dust configuration, the emitted SED
will depend on the properties of the central star,
such as the effective temperature and luminosity.
In the case of the cool class Flat objects
we showed that their de-reddened SEDs can look like the SEDs
of the hotter (and more massive) CTTS,
but when compared to other objects of the similar spectral type/mass
represented by the M-type IC348 templates of \citet{Lada06},
we observe that these Flat YSOs posses more emission
and thus can be less evolved than class II.
Similarly (see Appendix), we showed for small excesses,
by comparing SED fitting with our and with Winston's et al. (2009)
SpTs, that they can be interpreted either
as thinned ([K99]40), gapped (H232), or normal thick disks
depending on the amount of de-reddening and the selected SED template,
both of which are derived using SpT.
Our study also emphasizes the need to develop models of the circumstellar environment
for the low-mass objects to understand the evolutionary status of the
empirical SEDs of \citet{Lada06}.

We now want to test an age sequence
between SED classes of Serpens objects.
Despite a wide-accepted theory that these classes represent
an evolutionary sequence,
little evidence exists for the age trend in the literature.
The task is difficult due to the limited samples,
different techniques for placing different SED classes
on the HR diagram, and large observational errors.
As our HR diagrams show (Fig. \ref{hrBCAH}, \ref{hrDM}),
class 0/I EC89 is one of the youngest
sources, being $<$1 Myr old.
Class II objects are mostly 1 -- 5 Myr-old,
and diskless stars are mostly $>$3 Myr old.
This is in contrast with the studies of \citet{Winston09} and \citet{Muench07} who
did not find any difference in the age distributions between class II and III sources
in Serpens/NGC1333 and IC348, respectively, based on larger samples.
In the light of the latter result it is hard to understand the observed evolution
of the disk properties observed by other and these researches,
such as the amount of extinction, the shape of the excess, gas accretion,
dust crystallinity, etc. 
We agree with these and other surveys \citep[e.g., ][]{Flaherty08},
though, about the existence of the diskless objects even at the youngest ages,
indicating additional factors for disk clearing besides age.
Given the discussed uncertainties of placing and interpreting
YSOs on the HR diagram, we feel that more afford is needed in the direction of the
comprehensive study of the individual objects,
which we attempted in the the present work.

\subsection{Star Formation History in the Cloud}\label{sfh}

The HR diagram shows that the average age of the cluster is about 1 Myr.
A smaller distance of 259 pc
would shift the sources to fainter luminosities by 0.33 dex,
bringing the mean age to 3 Myr and increasing the age spread,
as indeed observed in \citet{Winston09}.
Six stars that could be potentially G-K cluster members
appear displaced on the HR diagram from the rest population.
Coupled with the lack of Xrays for the moderate reddening and high luminosity,
and the hints from spectroscopy (enhanced SiI or CO lines),
it points to the likely background giant nature.
On the other hand, 10 -- 20 Myr age is not totally excluded for class III members
in such clusters (see e.g. \citet{Luhman03, Winston09, Muench07}).
Given the lack of the classified intermediate-mass members in Serpens,
we list these objects in Table \ref{tableField} for the future investigation of their status
with a higher spectral resolution.

Our age estimate for the more certain members is similar to 
that inferred for the cluster from the luminosity functions in the literature,
and for the less rich cluster B further south in the cloud
\citep{Djupvik06}. The mean ages of 1 -- 3 Myr are typical
for embedded clusters. However, unlike some other studies,
we find only a moderate age spread of $\leq 5$ Myr
in the surveyed parsec-wide area around the cloud core,
source 19 would be the only member older than 10 Myr
according to both BCAH and DM tracks. 

In order to investigate the nature of the age spread,
one can examine the location of the various SED classes
in the cloud. Previous IR surveys already
identified that the core of the cluster
is richer in protostars, while the outskirts are dominated
by a much larger scattered population of T Tau stars \citep{Kaas04, Harvey06, Winston07}.
Furthermore, if we compare two
\textit{Spitzer}-selected YSO samples, one by \citet{Harvey07b}
that is limited to strong excesses, and another by \citet{Winston07}
with more relaxed excess criteria and inclusion of X-ray
sources, we see that the entire $0.5\degr \times 0.5\degr$
($\sim$3 pc wide) region surveyed in the latter study
is populated by active diskless stars, blurring
the cluster boundary.
But since the correlation between SED classes and ages
is not perfect, in Figure \ref{figIRACexc}
we show ages directly, as inferred from our BCAH98-based HR diagram.
One can see that objects younger than 1 Myr
are all concentrated in the core,
while the older ones lie further out.

One scenario suggests that the process of star formation in the
Serpens core is continuous, being responsible for both protostars
and the 1 -- 5 Myr-old dispersed population.
In our study however, there is an evidence
that not all YSOs in the halo are randomly distributed
(as would be expected if they migrated from the core).
Some YSOs delineate two dark filaments, one branching south and another
(less conspicuous, between H232 and [K99]40) east
of the main core, see Figures \ref{flam_big} and \ref{figIRACexc},
as well as Fig. 12 in \citet{Kaas04}.
Similar elongated orderings were found in many other young clusters
\citep{Allen07}, as well as in the Serpens South cluster \citep{Gutermuth08},
and imply multiple sites for star formation in the cloud.

Did individual groupings of YSOs form independently,
for example from the turbulent condensations in the cloud,
or was the whole process triggered by external force?
It has been proposed that winds and ionization fronts from
O -- B stars create a propagating wave of star formation
(e.g., \citet{Ikeda08,Reach08,Koenig08}).
In the case of the Serpens cloud, however, there are no nearby O stars.
There are only three late-B stars, including the illuminating source
of the S68 nebula, to the north-east of the core,
and a B9/A0 binary HD170545 to the south. 
The former stars may be responsible for heating the dusty ridge flanking
the Serpens cluster on the east side (bright area
seen on the left of Figure \ref{figIRACexc}), but it is not clear
how it could assist the cloud collapse.
A comparable or larger age spread than in Serpens has been found
in many other starforming regions without evidence of hot star intervention.
For example, \citet{Galfak08} 
describe a possible constant star-formation
over 50 Myr in the
low-mass region L1641N in Orion,
while \citet{Muench07} find a 5 Myr long episode in Perseus cluster IC 348.
Furthermore, \citet{Zuniga08} concluded for a rich Rosette complex
that the cluster ages cannot be satisfactory explained by
the propagation of the HII region.
Finally, it is impossible not to speculate on the role
that the intermediate-mass stars SVS2 and SVS20
may be playing in the cloud. They are centrally located
between two sub-cores and are surrounded by large cavities oriented SE-NW,
along the axis of protostars as seen in Figure {\ref{flam_big},
implying the same direction for the outflows.
\citet{Warren87} already proposed that the star formation
in the core could have been caused by the outflow of SVS2.
An outflow-triggered star formation episode was described
for example in the OMC-2 FIR 3/4 region in Orion \citep{Shimajiri08}
and in the Perseus cluster NGC 1333 \citep{Sandell01}.
The outflows may as well eventually destroy the cloud
and shut-down the star-formation process.

To differentiate among all these scenarios,
an age determination should be performed for a larger
number of cloud members. For example, Figure \ref{flam_zoom} shows that
there are several diskless YSOs in the vicinity
of SVS2 and SVS20. We investigated EC86, EC117, and EC84,
and found that formally their ages are $<$1 Myr,
being consistent with coeval origin with the protostars
in the region. 
It is vital to obtain ages for the rest of class III objects,
as well as to search for old stars
that would indicate a prolonged phase of star formation
instead of a burst.

\section{Summary and Conclusions}

We obtained $JHK$ images and $JH$ spectra in
a $10\arcmin \times 10\arcmin$ area centered on the Serpens reflection nebula
with an associated embedded cluster.
We determined SpT/$T_{eff}$ and other stellar properties
for 15 YSOs (plus 1 tentative) of SED classes 0/I through III,
in 6 cases they were derived for the first time.
Reddening for objects without NIR excess on the $JH-HK$ diagram
was estimated assuming a dwarf $JH$ intrinsic color,
while for objects with excess dereddening to the adjusted CTTS locus
was performed. 
We found that $A_{V}$ ranges from 2 to 24 mag,
precluding optical studies of some of these objects.
Spectral classification was performed against
the comparable spectra of YSOs in a similar-age cluster IC 348.
The SpTs turned out to be mostly early to late Ms,
unlike the late-K types of more massive CTTSs often assumed for class II objects
in the literature \citep{KH95, Meyer97}. The derived stellar parameters
allowed to construct an HR diagram and interpret the SEDs. 
The main results of this study are the following.
\begin{itemize}
\item According to the the BCAH98 theoretical tracks,
three objects are young brown dwarfs, all of the SED class II:
[K99]40, H230, and W68 (and possibly class I EC89).
Few more class II and III brown dwarf candidates are visible, but
but not discussed, on the HR diagram of \citet{Winston09}.  
Only one other brown dwarf
has been previously spectroscopically confirmed in Serpens,
and its SED class is unknown.

\item We critically evaluated previously published distance measurements
and concluded that 380 pc is a more appropriate value.
The earlier distance estimate of 259 pc
may be too low due to the adoption of the ZAMS,
rather than a dwarf, luminosity scale for field B -- A stars
for this particular region in \citet{Straizys96} and \citet{Straizys03}. 
One of the implications of the larger
distance scale is that a number of suspected PMS wide binaries
are unlikely to be gravitationally bound (such as EC 129).

\item Among 11 objects with NIR magnitudes comparable to
Serpens low-mass YSOs, eight are likely background reddened K -- M giants,
as indicated by prominent CO features in the $H$ band.
This technique only works for late-K to M types.
At earlier types it is extremely difficult to differentiate
YSOs without disks from old field stars.
X-rays, variability, and atomic line youth diagnostics in the optical spectra
(such as LiI and emission in H$\alpha$) can be used for the definite identification of class III YSOs.

\item The SEDs of the investigated objects are diverse:
from class I, CTTSs, evolved disks with holes,
to diskless objects.
We showed that the uncertain reddening in the SED class flat objects
can be a reason for assigning adjacent classes I and II
to them in the literature.
The de-reddened flat class SEDs of Serpens low-mass objects, however, do
differ from the SEDs of class II objects in Serpens and IC348
of a similar SpT, by having lager excesses at mid-IR.

\item We found two Fl/II objects $<$3 Myr-old (H232 and STGM1)
with on-going gas accretion
typical of T Tau stars as evidenced by emission in Pa$\beta$ .
H232 is especially interesting with depression in the SED short-ward
of 10 $\mu$m.  
We also detected He I emission line at 1.083 $\mu$m
in SVS2, the illuminating source of the Serpens
reflection nebula, confirming that this source drives an outflow or wind
and may be capable of destruction of the cloud.
A rich emission-line spectrum was found
in a protostar EC103, which is associated with a compact blue extension,
probably an outflow cavity.

\item Our data confirms that X-rays are good
indicators of YSOs, but not always for the more embedded class 0/I.
The sensitivity limit of {\it Chandra} survey for Serpens
is inferred to be at $logL_{star}/L_{\sun}\sim-1.5$,
roughly corresponding to a brown dwarf boundary at 1 Myr.

\item We determined the average age of the cluster to be about 1 Myr,
but with a range between $<$1 Myr in the cluster core
up to 5 -- 10 Myr in the corona. Unlike in \citet{Winston09}, we do not find
members beyond 30 Myrs, which makes it more similar with NGC 1333 from their study.
We found that class II objects to the east and south
of the core lie along two dusty filaments, seen in the NIR, {\it Spitzer}, and 850$\mu$m images of \citet{Davis99}.
Such filamentary structures are often observed in embedded clusters.

\item There is a tendency of age to correlate with the SED class:
class 0/I objects are $<$1 Myr,
and class III objects are on average older than classes Flat and II.
However, there is a number of extremely young sources
in the cluster core that appear to lack circumstellar
material (EC84, EC86, EC117). This remains true also
with the earlier SpTs of \citet{Winston09} for EC84 and EC86.
More sources like this can be seen in \citet{Winston09}.
Alternatively, these sources could be multiple systems or variables.
\end{itemize}

\acknowledgments

We would like to thank J. Chiar for making available for us
the chapter of her thesis on the distance to Serpens;
K. Luhman and J. Muzerolle for the discussion on the intrinsic colors of M stars;
the members of FLAMINGOS-I team
D. Joyce, N. Raines,  B. Ferreira, M. Franz, C. G\'{o}mez-Mart\'{i}n, M. Huerta,  J. Levine,
N. Rashkind, C. Roman-Zuniga, and A. Stolte
for assisting with observing runs and sharing the software;
J. Rossa and G. Rieke for proofreading the manuscript.
We acknowledge the anonymous referee for the critical comments
that helped improve the paper.
The data were collected under the NOAO Survey Program, {\it Towards a
Complete Near-Infrared Spectroscopic Survey of Giant Molecular
Clouds} (PI: E. Lada) and supported by NSF grants AST97-
3367 and AST02-02976 to the University of Florida. We also
acknowledge support from NASA grant LTSA NNG05D66G.
This publication makes use of data products from the {\it Spitzer}
Legacy ``Cores to Disk'' program and the Two Micron All Sky Survey.
Facilities: KPNO:2.1 m, Mayall, {\it Spitzer}, {\it CXO}.
\\
\\
\indent {\it We dedicate this work to Joern Rossa, our colleague and good friend at UF,
who passed away shortly after this paper was submitted.}
\\
\\
\appendix

\section{Comments on Individual YSOs}\label{who}

In this Appendix we discuss each of the candidate YSO in detail,
roughly in the order of decreasing
excess, as plotted on Figure \ref{figsed}.

{\bf EC89} is the only class 0/I object
that we could spectroscopically classify and place on the HR diagram
(the spectrum of another similar object, EC 103,
does not have sufficient S/N for spectral classification,
and in addition is veiled with emission lines).
It was assigned SED class 0/I in \citet{Winston07},
class I in \citet{Kaas04}, and not classified in \citet{Harvey07b}
except for a YSO designation.
It is difficult to measure the 24 $\mu$m flux
because it is a part of a dense mini-cluster of protostars, together with EC84.
\citet{Duchene07} consider it to be a component of a wide binary
system with class I EC88 (red protostar 4.4$\arcsec$ south
in Figure \ref{flam_zoom}, a radio source \citep{Eiroa05}). 
\citet{Duchene07} discovered that EC89 itself is a binary,
with a tight IR companion at 0.13$\arcsec$ separation,
corresponding to 49 AU for $d=$380pc.
We derive age of EC89 to be less than 1 Myr according to both sets of tracks,
and mass close to the brown dwarf boundary.
The 10 times larger mass and surprisingly old
age of 42 Myr deduced by \citet{Pontoppidan04} for this source ($=$SVS4-6)
must be due to their overestimation of $T_{eff}=$ 4750 K,
that corresponds to an early-K star as compared to our mid-M classification.
A $K$-band echelle spectrum is shown in \citet{Doppmann05},
with Br$\gamma$ in emission and CO band in absorption;
unfortunately, the authors were not able to derive
photospheric parameters due to low S/N. 
According to our estimates it is more embedded and intrinsically
more luminous than the comparable mass class II objects [K99]40 and H230.
After de-reddening the SED with $A_{V}=15.2$,
the excess remains higher than in the CTTSs at all wavelengths,
confirming {\it Spitzer} classification (Fig. \ref{figsed}).
EC89 was not detected in the X-rays \citep{Giardino07, Winston07},
as are none of class 0 objects in Serpens.
\\
\indent {\bf EC129} is intrinsically the most luminous object that we were able
to place on the HR diagram. Together with EC117 it is a solar-mass star
according to the BCAH98 tracks.
It illuminates a cometary nebula in the NIR images.
\citet{Kaas04} assigned SED class flat to EC129,
\citet{Harvey07b} class I, and \citet{Winston07} class 0/I.
Figure \ref{figsed} shows that the de-reddened SED lies
close to the CTTSs,
therefore, class II would be a more appropriate classification.
One can notice the lack of the NIR excess, which
can either be interpreted as an inner disk clearing
or if the star is of the slightly later SpT (M2-M3);
in the latter case, however, it will fall above the birthline
on the HR diagram.
We think that the assignment of the earlier SED classes
is due to the underestimation of the extinction --
the surveys of \citet{Kaas04}, \citet{Harvey07b},
and \citet{Winston07} do not report any value for $A_{V}$,
\citet{Eiroa92} obtained $A_{V}=16$ by adopting
constant $(H-K)_{0}=0.5$, while our
SpT-based method resulted in the larger $A_{V}=24$.
We speculate that the non-detection of X-rays may be also due to large extinction.
The $K$-band spectrum of
\citet{Doppmann05} exhibits weak absorption lines
that allowed them to estimate  $T_{eff}=4400 \pm 74$ K, $logg \leq 3.5$
based on comparison with NextGen spectra.
Br${\gamma}$ exhibits asymmetric double-peak emission.
The CO band at 2.29 $\mu$m is mainly in absorption,
but also contains a weak blue-shifted emission.
The spectrum is very similar to their spectrum of
GV Tau S, an evolved class I object.
In a more recent study \citet{Doppmann08} revised
$T_{eff}$ for GV Tau S from the original 4500 K to 3800 K,
modeling the absorption part of the CO profile
as arising from the  photosphere
and emission from the disk.
This implies that their estimate of $T_{eff}$ for EC129 could be also too high,
favoring our value of $3900\pm 300$ K.
\citet{Haisch04} detected a faint star 6.2$\arcsec$ north-west of EC129,
which can also be seen in our $K$-band image.
They speculate that it could be a wide companion to EC129.
However, with $d=380$ pc the separation would be 2360 AU,
beyond the the 2000 AU limit for the gravitationally bound systems they were looking for.
\\
\indent {\bf EC74} is classified as a flat class source
in \citet{Harvey07b} and \citet{Winston07}, with no $A_{V}$ reported,
but as  class II in \citet{Kaas04}.
Indeed, as seen on Fig. \ref{figsed},
the de-reddened SED is close to the CTTS one of \citet{Dalessio99}.
However, when comparing to the thick disk SED
of a mid-M source in IC348, we observe that the excess in EC74
is considerably larger.
According to the location on the
HR diagram, EC74 is $<$1 Myr and is likely to be in the embedded stage.
Absorption bands of silicates and ices have been detected in
the mid-IR spectrum of EC74 and other
YSOs in the Serpens core by \citet{Eiroa92}, \citet{Boogert08}, and \citet{Pont08}.
The most straightforward association of these species is
with the massive circumstellar envelopes.
This interpretation, however, is complicated by the possible
high-inclination disk effects \citep{Crapsi08},
and the contribution from the ambient material of the molecular cloud,
since similar bands are also observed in the background sources (like CK2).
\citet{Lahuis07} report the detection of H2 S(2),(3), and [NeII]
emission lines in the {\it Spitzer} spectra of EC74, indicative of a $\sim$800 K gaseous disk
within 2 AU from the star and of high-energy photons to heat it,
such as from the stellar chromosphere or accretion shocks. 
Indeed, EC74 is an X-ray source, unlike EC89 and EC129.
\citet{Kaas99} reported EC74 to be variable in the $K$ band on the time-scales
from 2 days to 1 year.
We classified this object using water bands only,
as our spectrum is too noisy for detection of atomic lines.
The low-resolution $K$-band
spectrum of \citet{Eiroa06} is mostly featureless, 
with a weak Br$\gamma$ in emission, which is expected from veiling
from a hot circumstellar material.
Whether the object is a CTTS or in a less evolved stage thus remains to be
further investigated.
It is intriguing that it might be situated
at the edge of the disk-like structure around SVS2 (Fig. \ref{flam_zoom}),
though the size of the latter ($>$10,000 AU) implies that it must be a shadow rather than
a physical structure (see discussion below).  
\\
\indent {\bf EC114} is another example of a low-mass
flat class YSO without an $A_{V}$ estimate in the {\it Spitzer}
papers. When de-reddened ($A_{V}=$10), the SED fits that of the CTTSs,
but rises above the IC348 disk templates for M stars.
It is an X-ray source.
\\
\indent {\bf H232} was classified as a flat SED source
in \citet{Winston07} without an $A_{V}$ estimate,
and as a CTTS (thick disk class II) in \citet{Harvey07b}
with $A_{V}=$8.6, which is close to our 7.3.
As seen from the $JH-HK$ diagram and from the SED,
the excess only starts at 3 $\mu$m and experiences an upturn at 6 $\mu$m.
It would still be consistent with a thick disk IC348 SED considering
its M type, except for the elevated excess at long wavelengths.
The contrast with the low NIR excess is remarkable,
reminiscent of transition objects, though to a lesser degree.
The source shows a prominent emission in Pa$\beta$ in our spectrum (EW$=$10.5$\pm$1 \AA\,),
indicating an on-going gas accretion.
Using de-reddened $J$-band magnitude to flux calibrate the spectrum,
we obtain the emission flux in the line
$log L_{Pa\beta}/L_{\sun}=-3.3 \pm 0.1 $.
Using the empirically calibrated dependence of \citet{Muzerolle98}
of the Pa$\beta$ luminosity on the accretion luminosity, we obtain
$log L_{acc}/L_{\sun}=-0.6 \pm 0.1 $, which is a typical value
observed in class II objects in Taurus and Ophiucus (see histogram in  \citet{Muzerolle98}).
\citet{Winston09} report also the presence of the CaII 0.8 $\mu$m triplet in emission in this star.
It is an X-ray emitter.

\citet{Winston09} obtained a slightly larger stellar luminosity (with our distance)
$log L/L_{\sun}=-0.05$ (vs. our -0.25 $\pm$ 0.11), but also an earlier SpT=M0 $\pm$ 1.5
(vs. our M3 $\pm$ 0.6), which still makes it a 1-2 Myr old low-mass star.
Earlier SpT means less reddening (we calculate $A_{V}$=6.0 mag vs. ours 7.3)
and a higher Teff for the model photosphere (3900 K vs 3400 K).
Both effects make the observed IR excess stronger,
but Lada et al's K6-M0 SED has also stronger excess than the M2-M3 one.
The net effect on the SED is that the object can still be classified as a CTTS.
Interestingly, the up-turn in the SED remains with the SpT of \citet{Winston09},
but it now manifests as the lack of the 3.5-5 $\mu$m emission compared to the template,
which in principle can be interpreted due to a gap in the disk.
\\
\indent {\bf STGM1} was characterized as ''likely class II'' in \citet{Kaas04} with $A_{V}=$6.7,
class flat in \citet{Winston07} with no $A_{V}$ estimate, and
CTTS (thick disk class II) in \citet{Harvey07b} with $A_{V}=$8.9,
compared to our 6.8 $\pm$1.
Similar to other discussed flat-class objects,
the excess deviates significantly from the thick disk
M-type templates and reaches levels of the solar-mass CTTS,
except perhaps at NIR wavelengths. 
It is another object beside H232 that shows Pa$\beta$ in emission
with EW$=14.5 \pm 2$ \AA. It translates to $log L_{Pa\beta}/L_{\sun}=-4.1 \pm 0.1$
after de-reddening, resulting in $log L_{acc}/L_{\sun}=-1.5 \pm 0.1$,
which also falls within a range of typical values
observed for class II objects in Taurus and Ophiuchus.
The accretion luminosities are comparable in H232 and STGM1 
when considering them relative to their photospheric luminosities:
$L_{acc}/L_{phot}=0.4-0.5$. Both objects have ages 1 -- 2 Myr.
We note that emission in Pa$\beta$ of the same level can be present
in other late-M Serpens objects as well, but it cannot be detected due to lower S/N of their spectra.
X-ray information is not available for STGM1 because
it is outside the {\it Chandra} FOV.
\\
\indent {\bf H230} and {\it W68} are the lowest mass
objects in our sample, young brown dwarfs with disks.
Unlike in flat objects, their SEDs are below that of the the CTTSs,
but fit nicely the thick disk late-M template in IC348.
W68 was classified as a class II object in \citet{Winston07},
but with very uncertain 2MASS photometry it did not pass the YSOs criteria
of \citet{Harvey07b}.
Winston et al's 2007 $A_{V}=5.8$ (as converted from $A_{K}$ using Cohen's law)
is somewhat larger than our $3.4 \pm 0.8$.
Winston's et al. 2009 optical SpT$=$M8.7 $\pm$ 5.0 is
consistent with ours M7.75 $\pm$ 0.75, while their (corrected to our distance) luminosity
is larger: $logL_{\star}/L_{\sun}=-1.64$ vs. -1.94 $\pm$ 0.09. The object is clearly substellar
and is younger than 1 Myr in both studies. \citet{Winston09}
from their optical spectrum find that it has an outstandingly low gravity that could indicate that it is
a background giant instead, but that measurement can be
erroneous due to the low S/N, given the large errorbars on their SpT.
H230 was assigned class II in both {\it Spitzer} studies.
\citet{Winston07} obtained  $A_{V}=4.2$, 
\citet{Harvey07b} $A_{V}=$7.7
and $logL_{\star}/L_{\sun}=-1.4$ (for our distance),
and we obtained intermediate $A_{V}= 5.6\pm 1.2$,  $logL_{\star}/L_{\sun}=-1.7\pm0.14$.
To remind, \citet{Winston07} de-reddened YSOs to the CTTS locus,
\citet{Harvey07b} to the $J-K$ color of a normal K7 dwarf,
and we to the lowered CTTS locus (as appropriate
for M dwarfs). 
Together with H232, [K99]40, and two more class II sources,
H230 appears to trace the southern edge of the dark lane that
branches eastward of the cloud core (Figure \ref{flam_big}).
The non-detection in the X-rays of H230 and W68 can be due to the low luminosity
of these objects, as well as the location on the
periphery of the {\it Chandra} field.
\\
\indent {\bf [K99]40} and {\it K407} present cases of disks with reduced emission,
usually called anemic or evolved and sometimes transition, though the latter
term is reserved for objects with thick outer disks.
In [K99]40 and K407 the excess starts at $\sim5$ $\mu$m
and stays below that of CTTSs. When comparing with IC348 M-types,
the excess in these YSOs is intermediate between the thick and the thin disk templates.
The ages ($\leq3$Myr) and the reddening values are comparable with the more ``normal'' class II objects;
for [K99]40 was obtained $A_{V}=3.1$ \citep{Winston07}, 7.5 \citep{Harvey07b},
and $8.4 \pm 1$ (this work);
for K407 $A_{V}=5.2$ \citep{Winston07}, 3.4 \citep{Harvey07b}, 2.1 \citep{Kaas04},
and $4.1 \pm 0.5$ (this work).
The values for the stellar luminosity are consistent
between (corrected for distance) \citet{Harvey07b} and us:
for [K99]40: $logL_{\star}/L_{\sun}=-1.2$ and $-1.3 \pm 0.15$;
for K407 $logL_{\star}/L_{\sun}=-0.8$ and $-0.8 \pm 0.1$, respectively.
\citet{Winston09} obtained a smaller luminosity for [K99]40:
-1.74, and a similar for K407: -0.74. But in combination with their later SpTs
(M9.0 $\pm$ 5.0 vs. our M6.75 $\pm$ 0.75 for [K99]40, 
M5.0 $\pm$ 1.0 vs. our M3 $\pm$ 0.5 for K407),
the interpretation remains -- [K99]40 is $<$1 Myr and may be substellar,
while K407 is older and more massive.
K407 is outside of the {\it Chandra} FOV but was detected by {\it XMM-Newton};
[K99]40 was not detected in the X-rays, but it may be
due to its low luminosity and high extinction.

How do SpTs of \citet{Winston09} fit the SEDs? For K407 the status
of evolved disk remains valid, as the difference in SpTs of 2 subtypes
results in the difference in reddening of only 0.3 mag. We only note that our
earlier SpT better describes its location on the $JH-HK$ diagram,
according to both FLAMINGOS and 2MASS photometry.
On the other hand, for [K99]40 later type and smaller de-reddening ($A_{V}=7.2$ vs. ours 8.4)
resulted in a good match with the thick disk M6$+$ template.
The comparison, however, is far less reliable than for K407 due to
the large uncertainty on Winston's et al. SpT and the lack of
NextGen models with the required $T_{eff}=$2400 K.
Adopting this late SpT would also move [K99]40 above the birthline.
\\
\indent {\bf SVS2} is the second (after SVS20) brightest IR source in the Serpens main core;
it is optically visible and illuminates the SRN.
It is a single star according to \citet{Duchene07}.
Despite the long history of study, however, there is no spectroscopic
estimate of the temperature and hence
the accurate estimate of the extinction, mass, and the evolutionary status.
These parameters have been constrained through various indirect methods,
such as: the visual brightness
\citep{Worden74}, the $H-K$ color \citep{Harvey84},
the temperature and the IR luminosity of the
gas \citep{Ungerechts84} and dust \citep{Nordh82}
in the cloud, the SED \citep{Wolfire87},
the depth of the 3.07 $\mu$m ice band \citep{Eiroa89},
and the brightness of the reflection nebula \citep{Pontoppidan05}. 
These considerations indicate SpT between B7 and F5,
$A_{V}=8-17$, and $L \sim 50L_{\sun}$, which translates
into a $\lesssim3 M_{\sun}$ star of a 2 -- 10 Myr age
using tracks of \citet{Siess00}.
The reason for the lack of a definitive
estimate of the spectroscopic temperature
is the optical faintness of the star ($V=17$)
and a weakness of the absorption spectrum.
The optical spectrum by \citet{Cohen79} shows a rising continuum with
only a weak emission in H$\alpha$ and possibly [OI] 6300\AA\,, and
was assigned uncertain ``M0:'' type.
The $K$-band spectrum of \citet{Doppmann05}
only exhibits emission in H2 and a double-peak weak emission in Br$\gamma$.
Similarly, our NIR spectrum is devoid of strong absorption lines,
and so we did not attempt to classify it.
However, we detect, to our knowledge for the first time,
an emission line that can be
identified with the HeI line at 1.0830 $\mu$m.
It is resolved with FWHM$\sim$500 km/s,
comparable to the width of the structured H$\alpha$ profile in \citet{Gomez97}.
Unfortunately, the line falls outside of our standard calibration window,
therefore an accurate flux estimate is not available from our observations.
This line is frequently observed in CTTSs.
According to the models of \citet{Kwan07},
a symmetric profile without a blue-shifted absorption
originates in a polar stellar wind
observed at inclinations $>70\degr$ (their Figure 7).
If strong wind is present in SVS2, it could be responsible
for the SRN cavity, indicating that
a single intermediate-mass star is capable of disrupting
its parental cloud and perhaps controlling
the rate of the low-mass star formation.

\citet{Kaas04} classified SED of SVS2 as flat,
\citet{Winston07} as class 0/I, both without reddening estimate;
\citet{Pontoppidan03} and \citet{Lahuis07} as a class II,
and \citet{Harvey07b} only marked it as YSO.
It clearly has a NIR excess (Fig. \ref{fig_jhhk}),
which may be responsible for line veiling.
\citet{Kaas99} reported variability
in the $K$ band on a yearly timescale.
Indeed, comparing our photometry
with \citet{Kaas99} (2MASS $JH$ magnitudes are saturated),
we observe that the difference is roughly along the reddening vector
corresponding to $A_{V}$ decrease by $4.5$ mag.
SVS2 shows signs of an outflow and a disk.
\citet{Huard97}, among others, from polarimetric observations
deduced that the star is surrounded by a cavity,
and even identified few H2 knots that could be excited
by the the active outflow of SVS2, though
the association is quite uncertain given the crowdedness in the region.
On the other hand, strong silicate emission features at 10 and 18 $\mu$m
observed in the SVS2 spectrum \citep{Kessler06, Boogert08}
are thought to originate in the surface layer of dusty
disks (e.g., \citet{Furlan06}).
Furthermore, \citet{Lahuis07} reported the detection of the H2 S(2)
emission line in the {\it Spitzer} spectrum, indicative of a $\sim$1000 K gas,
while \citet{Pontoppidan03} and \citet{Pontoppidan05} found evidence of 
hot CO gas in Keplerian disk within 1 AU from the star.
It is an X-ray source as are most of the CTTSs.
Finally, as seen in Figure \ref{flam_zoom}, SVS2 is surrounded by a dark lane
of diameter 1$\arcmin$--1.5$\arcmin$ (20,000--30,000 AU)
with a wedge angle $\sim$20$\degr$.
Such structures have been modeled as shadows cast by
disks with radius $\sim$100 AU inside the ISM cavities
\citep[e.g.,][]{Pontoppidan05,Stark07}.
To reconcile the strong silicate emission features characteristic of the face-on disk
with the edge-on geometry inferred from the shadow model,
the disk was proposed to be very tenuous,
while the shadow would be produced by the puffed-up inner rim,
which would be also responsible for the NIR excess.
Further study is needed to pin down the inclination
and understand the nature of this feeble disk
(e.g. whether it is in the stage of dispersal or growing).

Most recently, \citet{Winston09} obtained an optical
spectrum for this source, which they classified as K8 $\pm$ 1.5.
From the HR diagram it appears a modest 0.9 $M_{\odot}$, $\sim$2 Myr-old star,
which may be surprising considering the above properties.
The paradox can be resolved if the inferred luminosity
is strongly underestimated due to the edge-on viewing geometry.
Indeed, \citet{Pontoppidan05} report that they may have resolved the obscuring
central disk of SVS2($=$CK3) in their $H$ and $K$-band images, and predict a 30 times larger
luminosity, though with a hotter $T_{eff}=6000 K$ than $3900$ K assumed
by Winston's et al. SpT.
\\
\indent {\bf SVS20} is the brightest IR source in the Serpens main core.
\citet{Casali93} and \citet{Davis99} detected it also in the sub-mm (source SMM6),
and \citet{Smith99} and \citet{Eiroa05} at 3.5 cm (VLA 16),
but it was not detected in 
the 1.1 mm study of \citet{Enoch07}, perhaps due
to confusion in the larger beam.
However, as with SVS2, little is known about the star itself.
It is not observable in the optical, while the NIR spectrum is
almost featureless. The maximum value of $A_{V}=35$ was estimated by \citet{Harvey84}
from the $K$ magnitude assuming a B8 ZAMS excess-free star.
\citet{Kaas99} reported variability of 0.14 magnitudes in the $K$ band within a 1 year
time span. While our $K-$band measurement is saturated,
the photometry from the literature (2MASS,
\citet{Sogawa97}, \citet{Kaas99}) indicates
that SVS20 is located on the $JH-HK$ diagram just outside of the
reddened CTTS locus.
Clearly, the object possesses a strong NIR excess.
Indeed, the depth of the 3.07 $\mu$m ice feature indicates
considerably smaller $A_{V}=14-17$ \citep{Eiroa89} than is inferred
assuming excess-free colors.

\citet{Kaas04} and \citet{Winston07} classified SVS20 as a flat SED source,
and \citet{Harvey07b} as class I, all without an $A_{V}$ estimate.
Mid-IR spectra revealed deep ice features of H2O, CO, and CO2
\citep{Eiroa89, Alexander03, Pontoppidan03, Pont08}.
{\it Spitzer} spectrum overlaid on the SED
is shown in \citet{Boogert08}. It shows silicate emission
feature at 10 $\mu$m characteristic of a disk,
but much weaker than in SVS2 and double-peaked,
which must be due to superposition with absorption. 
SVS20 is clearly more embedded than SVS2,
but it is not clear whether it translates into an earlier evolutionary stage
until the contribution from the cloud is subtracted.

The complication arises also from the fact that SVS20
is a binary \citep{Eiroa87}.
The companion SVS20B ($=$SVS20N) is situated
1.7$\arcsec$ (650 AU for $d=380$ pc) north-east from the primary.
It is 1.5 mag fainter in the NIR than the primary,
but becomes of comparable brightness by 13 $\mu$m \citep{Ciardi05, Haisch06}.
In the {\it Spitzer} beams the components are unresolved.
\citet{Ciardi05} modeled their ground-based resolved mid-IR observations
with the following components:
1) a circumbinary envelope producing absorption bands
and half of the visual extinction;
2) SVS20A: $T_{eff}=7000-10,000$ K,
$L_{\star}=40-160 L_{\sun}$ (for our distance),
corresponding to an A -- early-F $\sim 3 M_{\sun}$ YSO,
with the SED slope $\alpha_{2-10}=-0.3 $,
which places it at the Flat/class II boundary;
3) SVS20B: $T_{eff}=2800-3800$ K,
$L_{\star}\sim2 L_{\sun}$,
corresponding to an M-type $0.1-0.5 M_{\sun}$ YSO,
with $\alpha_{2-10}=-0.1$ characteristic of a flat-class SED.
The $K$-band spectrum of SVS20A in \citet{Doppmann05}
shows weak absorption features with a hint of emission in Br$\gamma$ and CO;
they deduced $T_{eff}=$5700 -- 6100 K
and $L_{star}=120 L_{\sun}$,
corresponding to an early-G star.
The $K$-band spectrum of SVS20B, on the other hand,
reveals prominent emission in Br$\gamma$, NaI, and CO lines \citep{Eiroa06};
the $M$-band spectrum of \citet{Pontoppidan03}
in addition shows a blue-shifted emission component in CO,
interpreted as an outflow.
Furthermore, the cm emission matches the position of the secondary
better, and \citet{Eiroa05} speculate
that it can be due to a thermal jet. 
Both components are X-ray emitters.
The secondary thus appears in a less evolved
stage than a more massive primary.
Recently, \citet{Duchene07} detected a tertiary component
in the system, with $\Delta K=4$ at 0.32$\arcsec$ (120 AU) to the west of SVS20A.
As seen in Fig. \ref{flam_zoom},
SVS20 is surrounded by a spectacular
clumpy ring of NIR emission, with a radius $\sim$10$\arcsec$ (3800 AU).
It has been speculated to represent either a circumbinary disk
or walls of a bipolar cavity, but no detailed modeling has
been performed yet.

Summarizing, SVS2 and SVS20 present examples of how auxiliary information
available for bright sources helps breaking degeneracies associated
with interpretation of the SEDs. 
These most luminous YSOs in the cloud with class I/Flat SEDs
appear to be embedded intermediate-mass stars with disks
and perhaps circumstellar envelopes.
Lacking ionizing radiation of B stars,
they may still be capable of affecting star-formation in the cloud
through the outflows/winds, judging from
the spectroscopic signatures and the large cavities surrounding
these sources.
\\
\indent {\bf EC103} is another class 0/I object
with unknown stellar properties.
It is variable on the timescale of 2 days \citep{Kaas99},
is not detected in the X-rays despite being a relatively bright
IR source, and has the strongest NIR excess judging by
location on the $JH-HK$ diagram.
Figure \ref{flam_zoom} shows a chain of Herbig-Haro (HH) objects
south of EC103. \citet{Herbst97} proposed that either class II YSO
CK8 or a deeply embedded sub-mm source SMM3 situated
10$\arcsec$ south, are the driving sources
of jets that produce these knots of shocked gas.
An examination of our image, however, shows
an elongated blue structure emanating south from EC103
in the direction of the knots.
Our spectrum of EC103 shows a steeply rising continuum
indicating $A_{V} \sim 20^{m}$
with the Brackett series in emission,
similar to a class I object in Taurus IRAS 04239+2436 presented by \citet{GL96}.
It also resembles spectra of Flat / class II sources
with high accretion rates from the latter study,
like DG Tau, HL Tau, and WL 18, where one can additionally see emission
in Pa${\beta}$.
All three sources are known to drive jets
\citep{Eisloffel98, Anglada07, Gomez03}.
The high-resolution $K$--band spectrum of EC103 in \citet{Doppmann05}
is similar to that of SVS2 (mostly featureless with narrow H2 emission),
except for a much stronger Br$\gamma$ emission.
On the other hand, the $K$-band spectrum of CK8 ($=$EC105)
from \citet{Eiroa06} does not show any emission lines.
We therefore consider EC 103 as an interesting alternative
outflow candidate to CK8 and SMM3.
The slight apparent misalignment of the blue extension
of EC103 and the HH chain may result from the uneven illumination of the cavity,
similar to the HH92 object studied by \citet{Bally02},
while the northern lobe can be too extincted to see.
\\
\indent {\bf EC117}, together with EC86 and EC84,
presents a puzzling case of an apparently very young star ($<$1 Myr)
without circumstellar material (SED class III).
It was proposed to be a YSO
based on the association with a faint cometary nebulosity
and later on the (variable) X-ray emission.
The low-resolution $K$-band spectrum of \citet{Eiroa06}
shows Na I, Ca I, and CO lines in absorption,
consistent with our early-M classification and with
the M0.5 $\pm$ 1.5 type of \citet{Winston09}.
The reddening estimates in the literature are even higher than ours ($A_{V}=$13.8):
15.1 mag in \citet{Kaas04},
18.7 mag in \citet{Winston07},
which would make it only more luminous on the HR diagram
(indeed \citet{Winston09} obtains (corrected to our distance) $logL/L_{\odot}=0.74$ vs. ours 0.51 $\pm$ 0.09).
Interesting, it is the third brightest
centimeter source in the central $2\arcmin \times 2\arcmin$ area
as revealed by VLA \citep{Smith99, Eiroa05}, outshining even SVS20.
\citet{Eiroa05} speculate that the radio emission in EC117 can originate
from the stellar corona. It is unclear at present
whether strong magnetic activity and the elevated location
on the HR diagram have a common cause, such as
binarity or extreme youth.

\indent {\bf W201} and {\bf EC77} are class III sources. 
W201 was selected as a YSO based purely on X-ray emission,
while EC77 based on X-rays and in the earlier works on the weak NIR excess
(which is not seen in FLAMINGOS or 2MASS photometry). They are likely diskless objects,
as indicated by the lack of excess through {\it Spitzer} 3 -- 8 $\mu$m bands,
and hence were not included in the YSO list of \citet{Harvey07b}.
EC77 is adjacent to the SVS4 mini-cluster, while W201 is situated
further south in the area of lower extinction.
Indeed, we obtain for EC77 $A_{V}=4.5 \pm 0.6$ (which compares well
to 3.2 mag of \citet{Winston07}), while for W201 we obtain smaller $A_{V}=2.1 \pm 0.5$.

\citet{Winston09} obtained a similar SpT to ours for W201 (M5 $\pm$ 1)
and later, but much more uncertain, for EC77 (M8.5 $\pm$ 5.0 vs. our M5.1 $\pm$ 0.4).
The distance-corrected $logL$ of \citet{Winston09}
-1.27 for W201 and -1.54 for EC77 are comparable with ours:
-1.12 $\pm$ 0.07 and -1.35 $\pm$ 0.07 respectively.
What effect would the later SpT of EC77 have on its SED?
Using SpT of \citet{Winston09}, we obtain $A_{V}=3.4$
and a 2600 K for the photospheric model, compared with ours
4.5 and 3100 K.
We obtain an equally good SED fit with these new parameters
as with ours.
This is because the cooler photospheric model is compensated by
a smaller de-reddening.
According to our HR diagram, the two stars represent $\sim$0.2 $M_{\sun}$ stars 3 -- 5 Myr old.
The lack of the circumstellar material is therefore not surprising,
as majority of stars dissipate
their optically-thick disks by this age \citep[e.g.,][]{Hernandez08}.
Less certain type of \citet{Winston09}, on the other hand,
would make EC77 much younger, $<$1 Myr.

Stars {\bf 19} and {\bf 28} are the only M dwarfs in our
spectroscopic sample that have not been previously reported as
cloud members. They do not posses IR excess
and are not detected in the X-rays, but the latter can be
due to low luminosities. On the HR diagram
they are markedly older than the other Serpens YSOs,
source 28 would be $\sim$30 Myr-old, while source 19
would be 10 -- 15 Myr-old according to the BCAH98 tracks.
The two sources are located next to the candidate YSO W201
(and the massive class III binary HD 170545),
therefore one can expect them to have similar reddening
if they belong to the cloud.
For W201 we obtained $A_{V}=2.10 \pm 0.54$,
and similar value for source 19: $2.45 \pm 0.55$,
while for source 28 it is only $1.01 \pm 0.57$.
Using theoretical luminosities of field dwarfs from BCAH98
and empirical from \citet{Leggett92, Leggett02},
and considering the $A_{V}-d$ diagram in \citet{Straizys03}
(except for ``sub-area A'' where distances may be underestimated),
we conclude that star 28 could be easily explained as a forground
dwarf at distances 250 -- 300 pc. Star 19, on the other hand,
is more difficult to explain as an interloper. Even if it is as far as 230 pc,
corresponding to the luminosity of a ``young disk'' population,
$A_{V}=2$ is quite rare there.
We conclude that star 19 remains a valid candidate
for a 10 -- 20 Myr-old 0.1 $M_{\odot}$ cloud member,
to be confirmed with more methods.

{\bf EC84} is a peculiar object in two respects.
Together with EC86 its location on the $JH-HK$ diagram is too blue
for its mid-M SpT, while IRAC fluxes fall below model SED.
According to \citet{Pontoppidan04},
it could be a K5 dwarf ($=$SVS4-4) with $A_{V}=$19.0,
or a K giant with $A_{V}=$17.5 -- 16.
Both explanations would contradict our spectra
as we do not see lines characteristic for K stars in the $H$ band,
while the giant explanation in addition would be inconsistent with
the X-ray emission from these sources \citep{Haisch92}.
We note that \citet{Eiroa06} show a low-resolution $K$ band spectrum
of EC84 with Na I and CO lines in absorption,
with strengths consistent with an M-type classification.
 
It is also extremely bright/young on the HR diagram
for a source with little to no IR excess.
According to Flamingos, 2MASS, and photometry of \citet{Eiroa92},
it does not have a NIR excess and was originally identified
as a YSO based on polarization and association with the SVS4 group
of YSOs \citep{Sogawa97}.
It appears to have a weak excess beyond 8 $\mu$m.
\citet{Kaas04} and \citet{Winston07}  classified it as class II object,
while \citet{Harvey07b} did not identify it as a YSO at all.
The long-wavelength fluxes are not
reliable because of the proximity to the bright proto-stellar EC92/95 system.
If excess is confirmed, it would be the thinnest disk in our survey,
perhaps even of a debris nature.
Other researches obtained similarly
high values of $A_{V}\sim20$ and (distance-corrected) $log L/L_{\sun} \sim 0.2-0.4$
\citep{Kaas04, Winston07, Winston09}.
The membership in the cloud for this object is essentially based on
the the fact that it is an X-ray emitter, and in addition
a strong 3.5 cm source, similar to another ``over-luminous''
class III source EC117.
\citet{Leggett06} cites N. Cross and M. Connelley
(private communication) that EC84 is a possible binary,
although decreasing luminosity by 0.30 dex
would still make it appear exceptionally young on the HR diagram.

The best explanation for the SED discrepancy would be
an intermediate M3 $\pm$ 1.5 type of \citet{Winston09},
that provides a good fit to fluxes up to 6 $\mu$m with a purely photospheric emission
reddened by $A_{V}=18.9$. The new SpT also moves EC84 below
the birthline on the HR diagram, but it still appears very young.
It remains a puzzle why it lost its optically-thick disk
at an age $<$0.1 Myr. 

{\bf EC86} is another example of a $<$1 Myr-old
star without circumstellar material.
It was selected as a YSO based on proximity to SVS2, the ice feature ($\tau=0.13$,
\citet{Eiroa92}), and X-rays.
Similar to EC84, our SpT-based $A_{V}=10$ is too large
to be compatible with the SED, and so is $A_{V}=12$
of \citet{Winston07}.
We are convinced that the star is of M type even though the spectrum was observed
at the edge of the array and continuum in the J band is affected
by a cross-talk depression. If it were a K-type star
we should have seen absorption lines in the $H$ band,
since veiling is excluded due to lack of the NIR excess
(as seen from the location on the $JH-HK$ diagram).

\citet{Winston09} classified it as M2 $\pm$ 1.5 from the NIR spectrum
(their late-M optical type should be discarded due to low S/N,
according to the private communication with E. Winston).
This early-M type with $A_{V}=9.0$ provides a very good fit
for the observed SED. The object remains $\lesssim$1 Myr
on the HR diagram. It maybe that the wind from the nearby SVS2
played a role in the early dispersion of disk around EC86.

Star {\bf 26} is a potential intermediate mass cluster member,
since on the HR diagram the errorbars
allow it to be younger than 3 Myr. 
However, the {\it Spitzer} fluxes could not be fit
with the $T_{eff}$ and $A_{V}$ inferred from our SpT and
the assumption of dwarf colors.  
The G dwarf classification is based on the
enhanced Si I 1.59$\mu$ line compared to the neighboring Mg I.
By looking at Figure 5 of \citet{Ivanov04}, however, we see that
similar spectra are observed in the metal-poor
K giants. If this star is a background giant, it will explain
the smaller $A_{V}$ required to fit the SED and the lack of X-rays.
Thus in the present study we do not consider it a cluster member.
The example of source 26 and the lack of definite G -- K members in our sample
shows how challenging is the identification
of the intermediate-mass members in young clusters
near the galactic plane.

\clearpage
\begin{figure}
%\epsscale{0.9}
\plotone{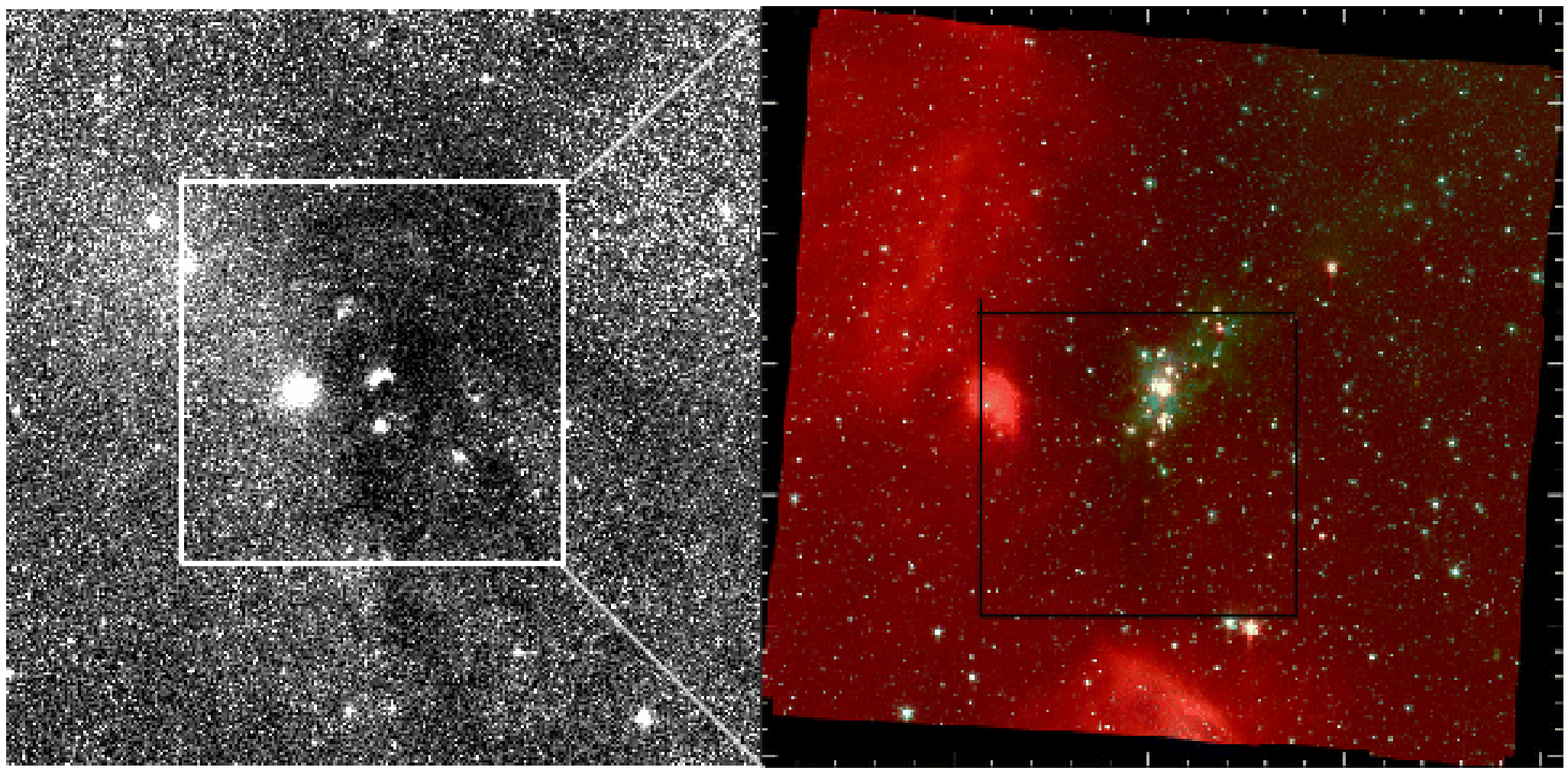}
\caption{{\it Left panel:} DSS $1 \degr \times 1\degr$ image 
of the area surrounding the Main core of the Serpens molecular cloud.
The butterfly-shaped nebula in the center
is the Serpens reflection nebula,
illuminated by the protostar SVS2.
Bright patch to the east is the reflection nebula Sharpless 68 (S68).
Like the two other nebulae to the north-east,
it is illuminated by a field late-type B star.
These stars were used in the literature for distance determination.
{\it White square / right panel:} {\it Spitzer}
image of the cloud core,
in 4.5 $\mu$m (green) and 8.0 $\mu$m (red)
bands, from \citet{Winston07}.
The brightest source in the center is SVS20A.
Black square: $10\arcmin \times 10\arcmin$ area spectroscopically surveyed
in this work (see Figure \ref{flam_big}).
North is up, east is left.
}\label{dss_irac} 
\end{figure}  

\clearpage
\begin{figure}
%\epsscale{0.9}
\plotone{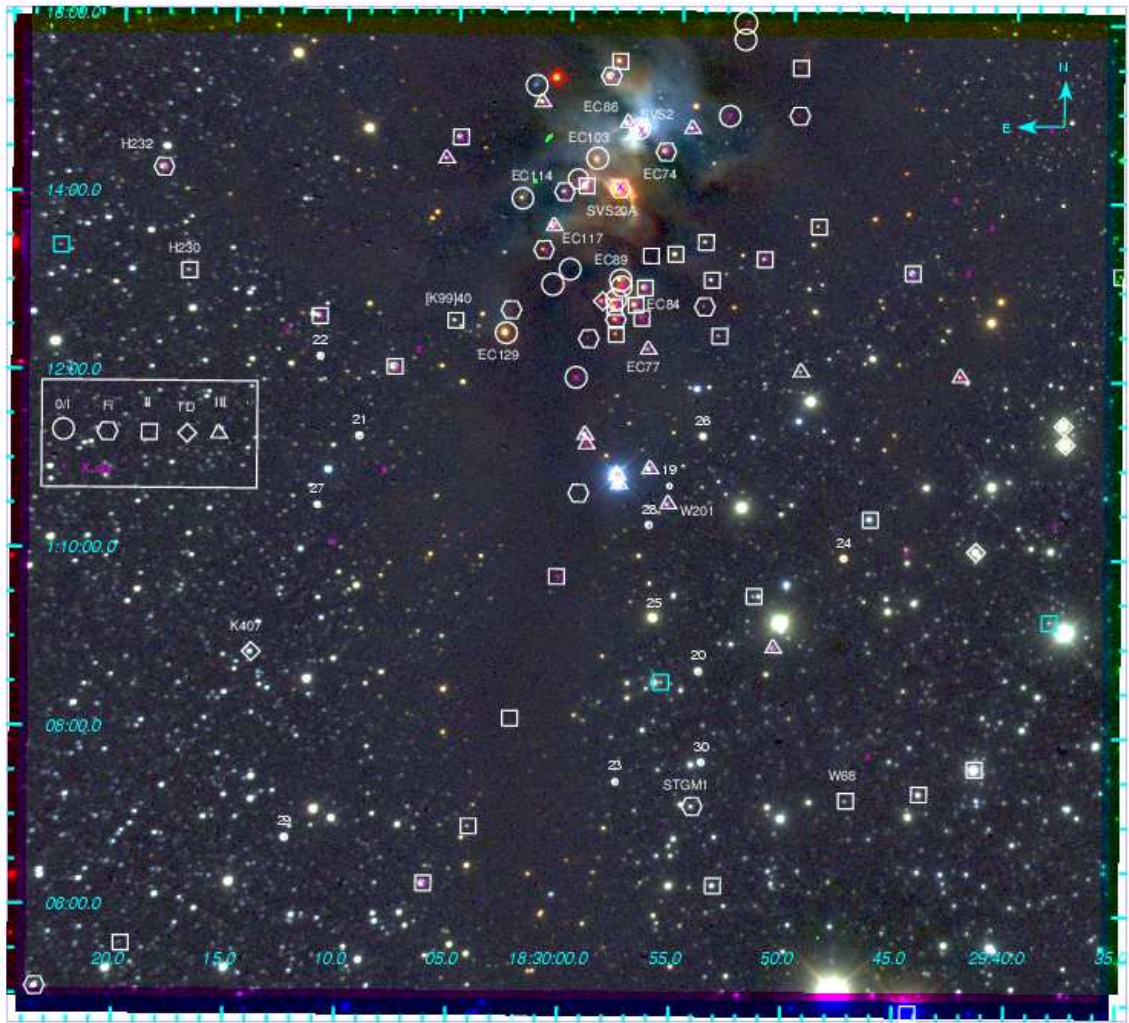}
\caption{FLAMINGOS $JHK$ image (square root flux stretch)
of our spectroscopic field, a subsection of the
$20 \arcmin \times 20 \arcmin$ image obtained at 2.1 m KPNO telescope.
Dimensions: $10 \arcmin \times 10 \arcmin$ on the side or 1.1 pc for d$=$380 pc.
YSOs from the {\it Spitzer$+$Chandra} study of \citet{Winston07} are shown
with different {\it white symbols} corresponding to different SED classes.
{\it Cyan squares} are some of the Winston's et al. (2007) sources that have not been regarded
as YSOs in the {\it Spitzer} study of \citet{Harvey07b},
likely due to faintness in 2MASS and {\it Spitzer} bands
(W56, W58, W64).
{\it Chandra} detections from \citet{Giardino07} are shown as {\it red crosses}.
Labeled are the spectroscopic targets considered in our paper.
}\label{flam_big} 
\end{figure}  

\clearpage
\begin{figure}
%\epsscale{0.9}
\plotone{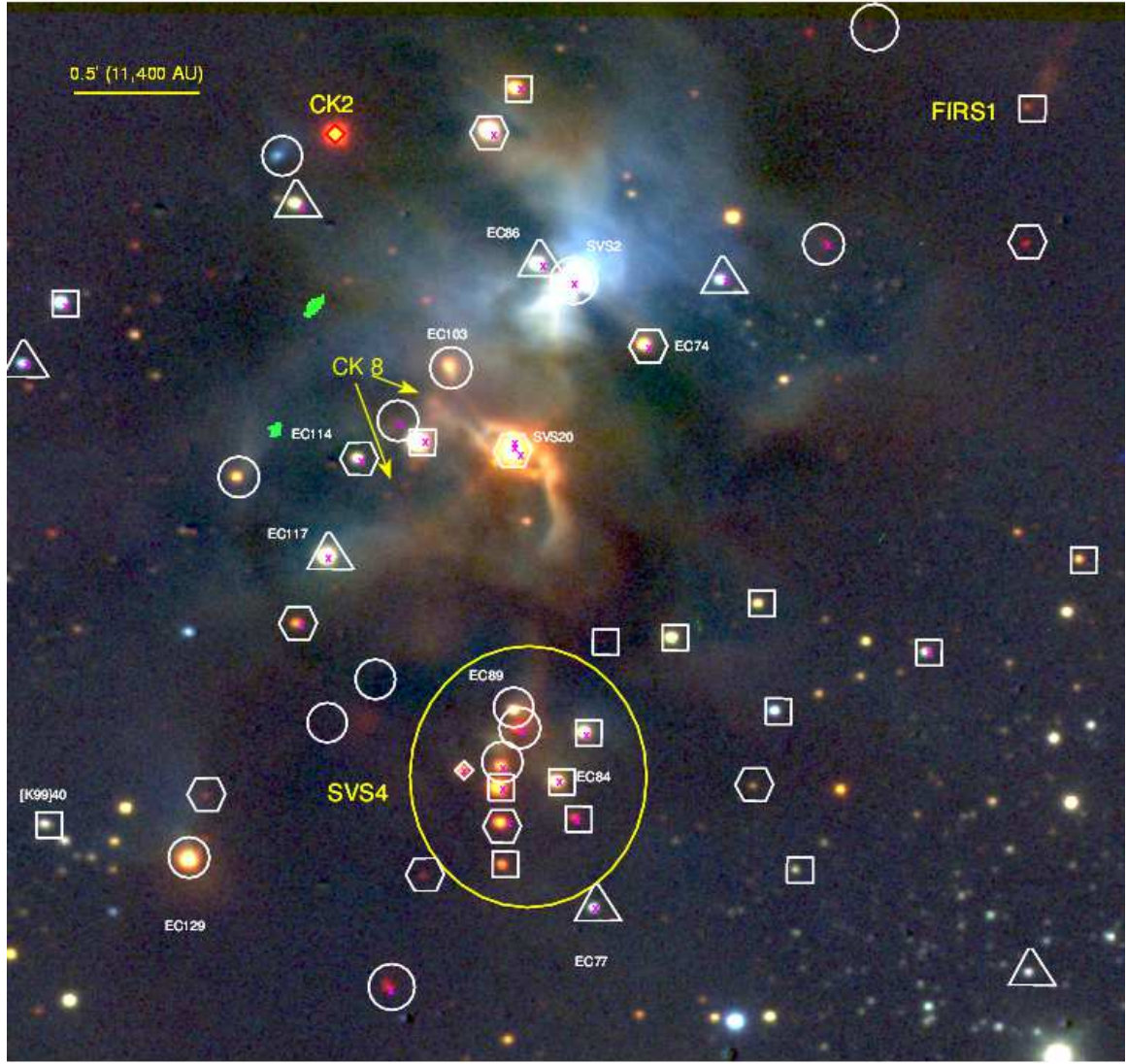}
\caption{A $4 \arcmin \times 4\arcmin$ zoom-in on the Serpens reflection nebula.
Symbols are the same as on the previous figure. A few notable objects are marked 
in addition to our spectroscopic targets for comparison with other studies:
SVS4 -- a compact cluster of protostars;
CK8 (EC105) -- class II source surrounded by a chain of Herbig-Haro objects;
FIRS1 (SMM1) -- a strong far-IR/radio source driving outflow;
CK2 -- classified as a YSO with a transition disk in \citet{Winston07}
(W158), but shown to be a background supergiant in \citet{Casali96},
which agrees with the c2d classification
of an $A_{V}=46$ reddened non-excess star.
The two green patches between CK2 and EC 114 are residuals of the sky
subtraction in the H band.
A SE-NW jet feature from SVS20 is actually a diffraction spike.
}\label{flam_zoom} 
\end{figure}  

\clearpage
\begin{figure}
%\epsscale{0.9}
\plotone{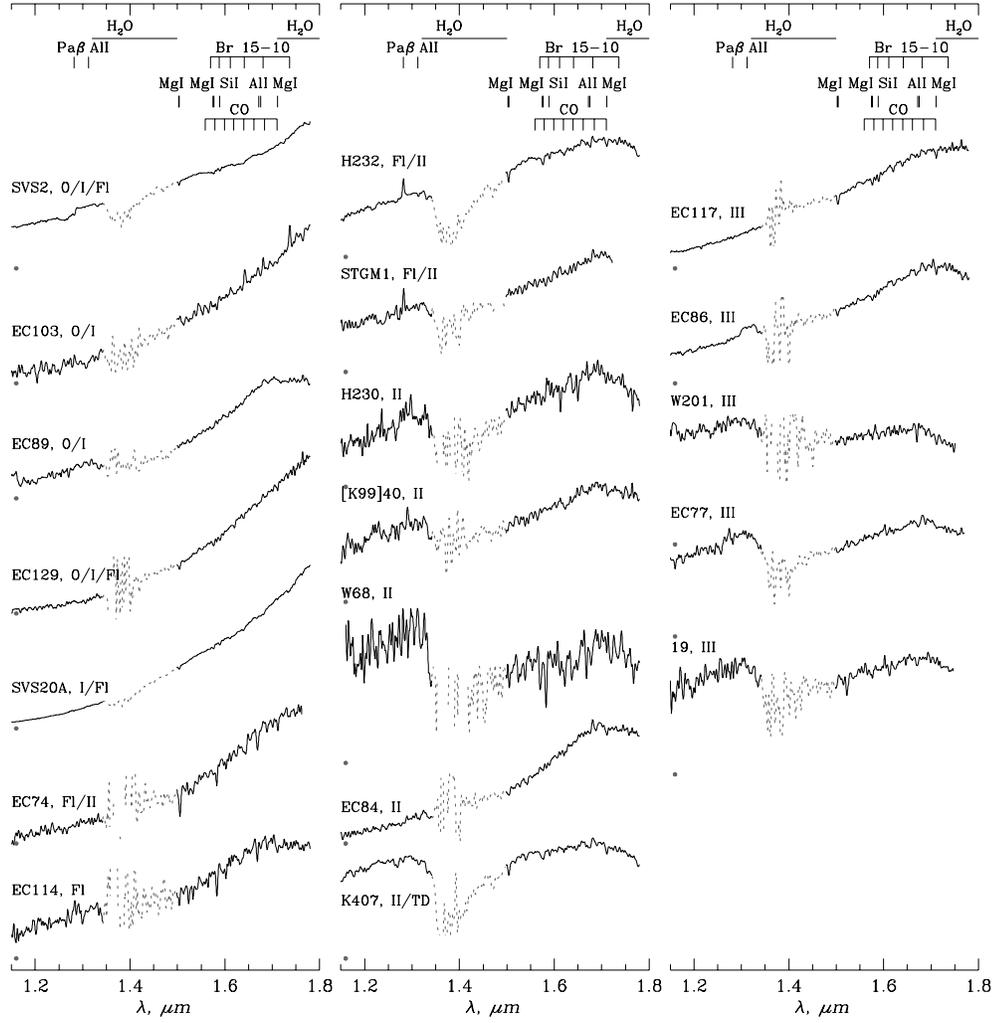}
\caption{Observed FLAMINGOS $JH$ spectra of YSO candidates from Table \ref{tableYSO},
plotted in three columns according to the \textit{Spitzer} SED class.
The spectra of protostars are mostly featureless with a steeply rising continuum,
while Class II and III objects reveal water bands and in some cases atomic lines.
Three objects show hydrogen emission lines: EC103, H232, STGM1.
The spectra have been normalized at 1.68 $\mu$m and shifted
along the y-axis; the zero flux level for each spectrum is shown as a gray dot.
The dashed segment at 1.4 $\mu$m is a noisy
region with telluric water residuals.
Resolution is $R=500$. The individual lines are indicated at the top of the figure.
}\label{fig_sp_under_yso} 
\end{figure}  

\clearpage
\begin{figure}
%\epsscale{0.9}
\plotone{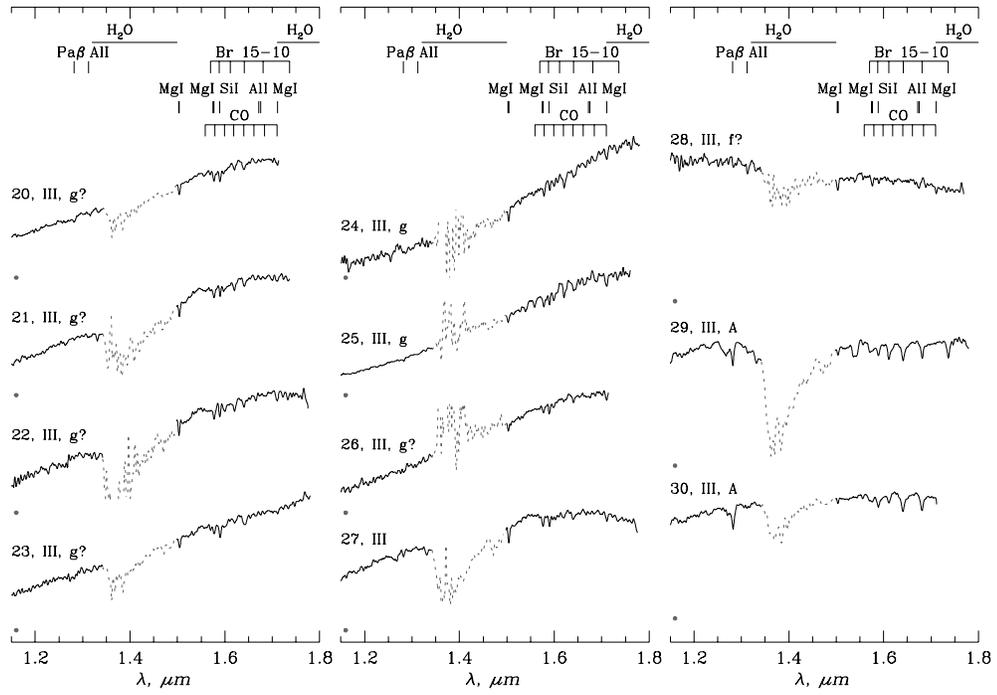}
\caption{Observed FLAMINGOS spectra
of the field star candidates from Table \ref{tableField}.
}\label{fig_sp_under_fld}   
\end{figure}  

\clearpage
\begin{figure}
%\epsscale{0.9}
\plotone{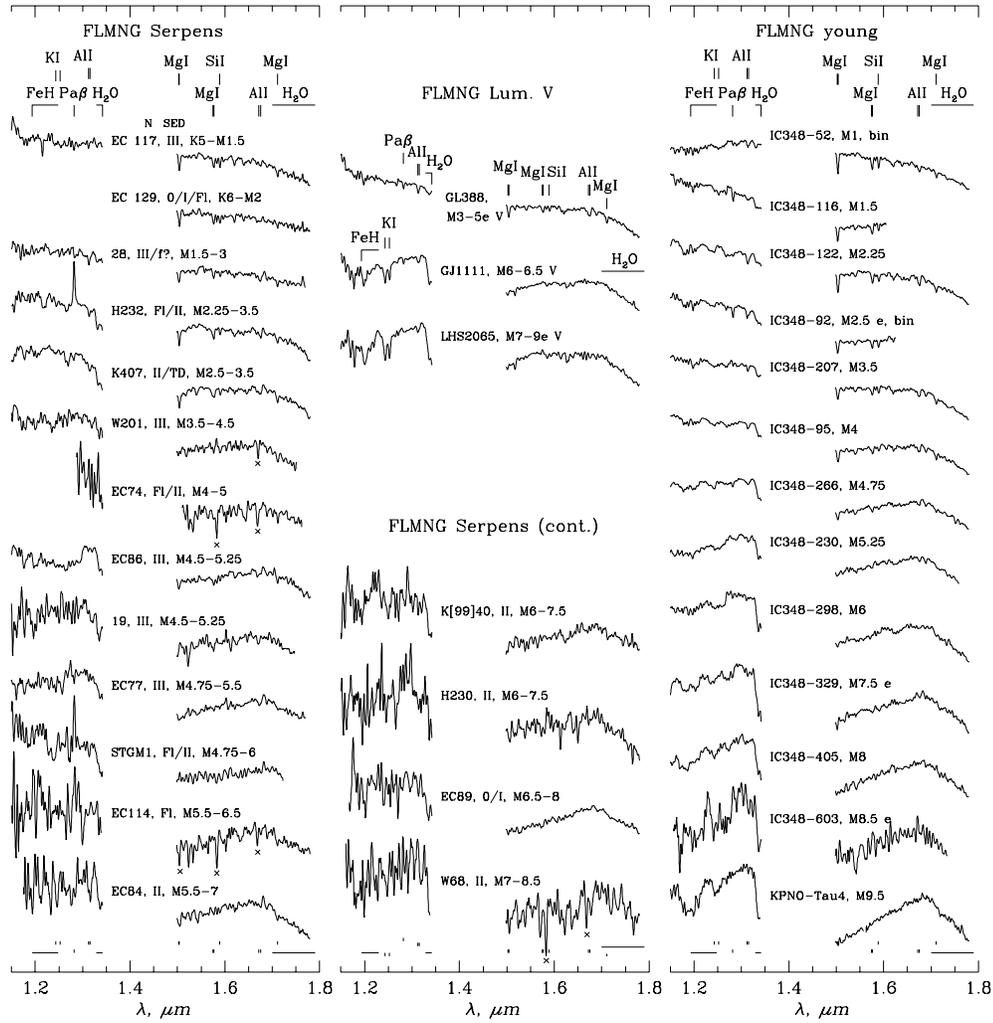}
\caption{FLAMINGOS spectra of M-type Serpens stars
compared to the optically-classified members of the 1 -- 3 Myr old
clusters IC 348 and Taurus-Auriga, and the field dwarfs (luminosity class V). To facilitate
a comparison for spectral classification of Serpens stars,
all spectra have been normalized to unity at 1.68 $\mu$m and dereddened to have flux
$\sim$1.2 at 1.32 $\mu$m.
Crosses in some spectra mark residuals from the three strong telluric OH lines.
Line identification: \citet{Origlia93, Meyer98, Cushing05}.
}\label{figMs}
\end{figure}

\clearpage
\begin{figure}
%\epsscale{0.9}
\plotone{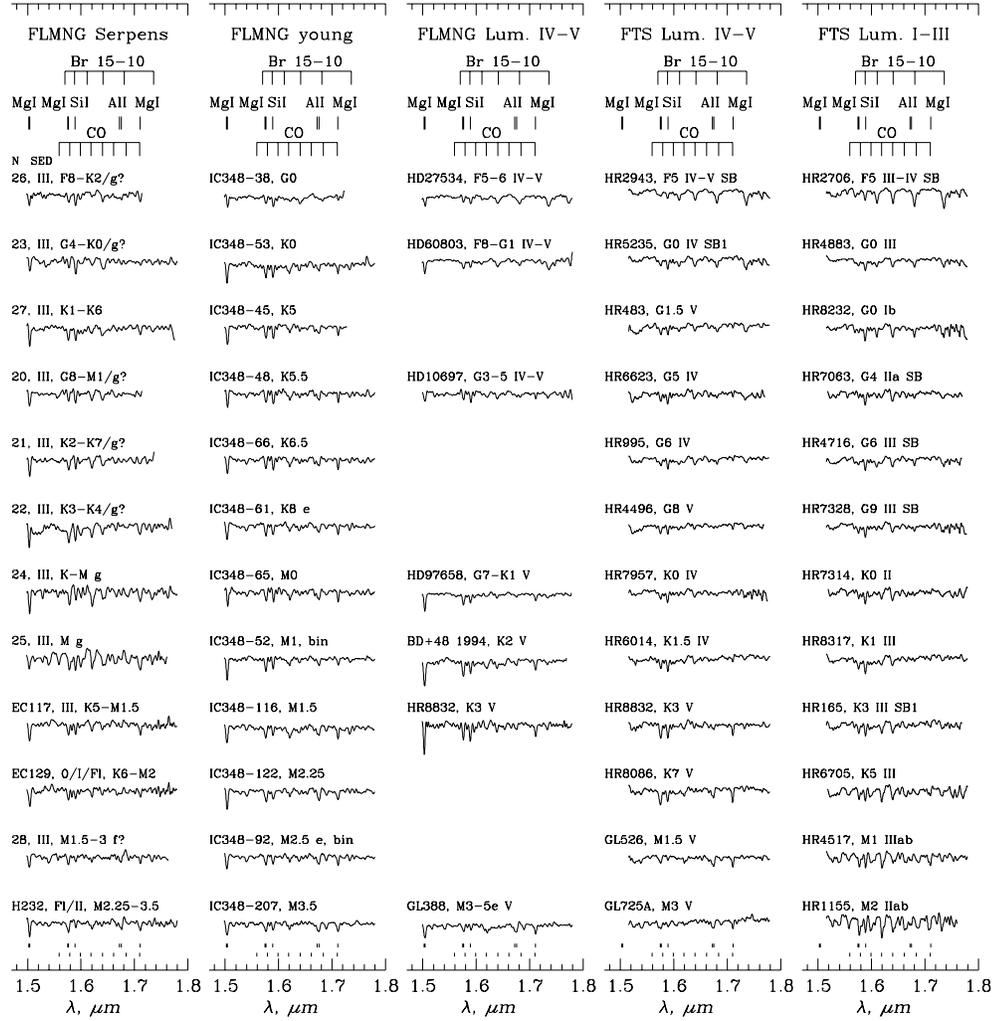}
\caption{Classification of Serpens spectra without water bands.
First column: FLAMINGOS $H$-band spectra of Serpens stars
(not shown: SVS2, EC103, SVS20A, 29, 30). Columns 2 and 3: FLAMINGOS
spectra of IC 348 early type members with classification
from \citet{Luhman03}, and of field dwarfs and subgiants with classification from VizieR.
Last two columns: FTS spectra of field stars from \citet{Meyer98},
separately for dwarfs/subgiants and low-gravity giants/supergiants.
Designations: {\it f?} -- candidate foreground field dwarf,
{\it g} -- strong field giant candidate, {\it g?} -- possible field giant.
Spectral types for g? group are
not reliable, as they are based on the comparison of the
atomic line strengths to that of IC 348 YSOs, solely for the purpose
of the placement on the HR diagram.
All spectra have been continuum-flattened and convolved to R$=$500.
Tick-marks for metal and CO lines are repeated at the bottom.
}\label{figGKs}
\end{figure}

\clearpage
\begin{figure}
\epsscale{1.0}
\plotone{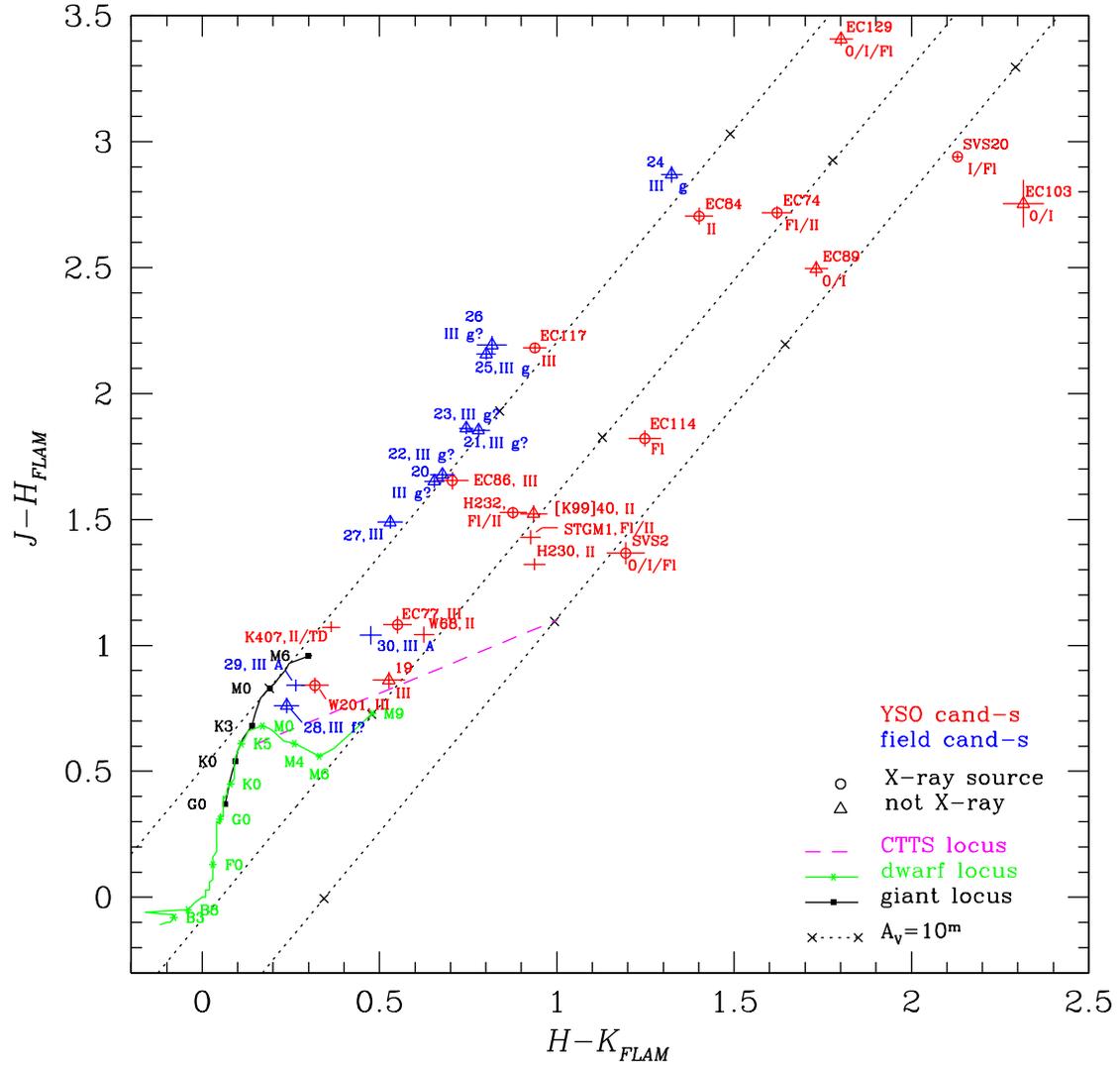}
\caption{$JH$ vs. $HK$ diagram for our spectroscopic targets in Serpens
(with SED classes marked).
Photometry for SVS20 is from \citet{Kaas99}.
\textit{Red symbols:} YSO candidates,
\textit{Blue symbols:} field star candidates;
\textit{g:} giants or supergiants based on our $H$ band spectra;
\textit{f:} probably a foreground star; \textit{A:} B -- A stars of unknown status. 
\textit{Circles:} {\it Chandra} detections, \textit{triangles:} {\it Chandra} non-detections,
\textit{rest:} outside of {\it Chandra} field of view.
\textit{Lines}: loci of dwarfs (green), giants (black), CTTSs (magenta), and $A_{V}=10^{m}$
reddening vectors (dotted), as described in the text.
For objects that deredden on this diagram
to the dwarf locus at the expected or earlier SpT,
$A_{V}$ is determined from $J-H$ color and SpT-$(J-H)_{0}$ dwarf calibration.
The remainder objects are dereddened to intercept
the adjusted CTTS locus, as explained in \S \ref{dered}.
X-ray non-emitters cluster along the uppermost reddening
vector, consistent with being field giants.
}\label{fig_jhhk}
\end{figure}  

\clearpage
\begin{figure}
%\epsscale{0.9}
\plotone{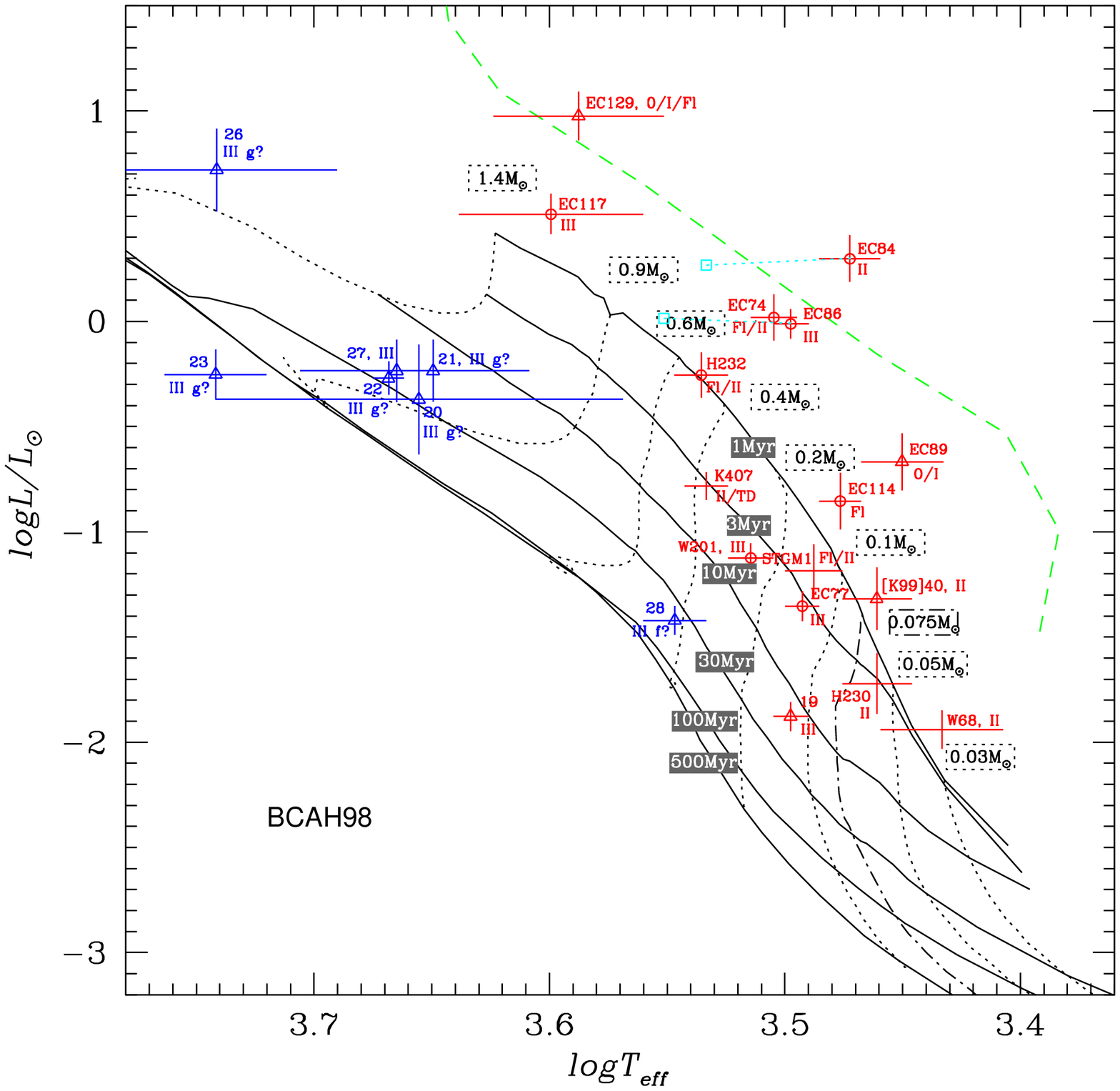}
\caption{Serpens sources overplotted on the evolutionary tracks
of \citet{bcah98} ($\alpha_{mix}=1.0$), assuming d$=$380 pc. Same symbols as in Figure \ref{fig_jhhk}.
Squares indicate positions of EC84, EC86 with the IR SpTs of \citet{Winston09}.
The brown dwarf boundary is marked with a dash-dotted line.
Dashed line designates a Deuterium-ignition birthline from \citet{DM97}.
One can see that the average age of the cluster is $\sim$1 Myr,
with class III objects being on average older than class 0--II.
}\label{hrBCAH}
\end{figure}  

\clearpage
\begin{figure}
%\epsscale{0.9}
\plotone{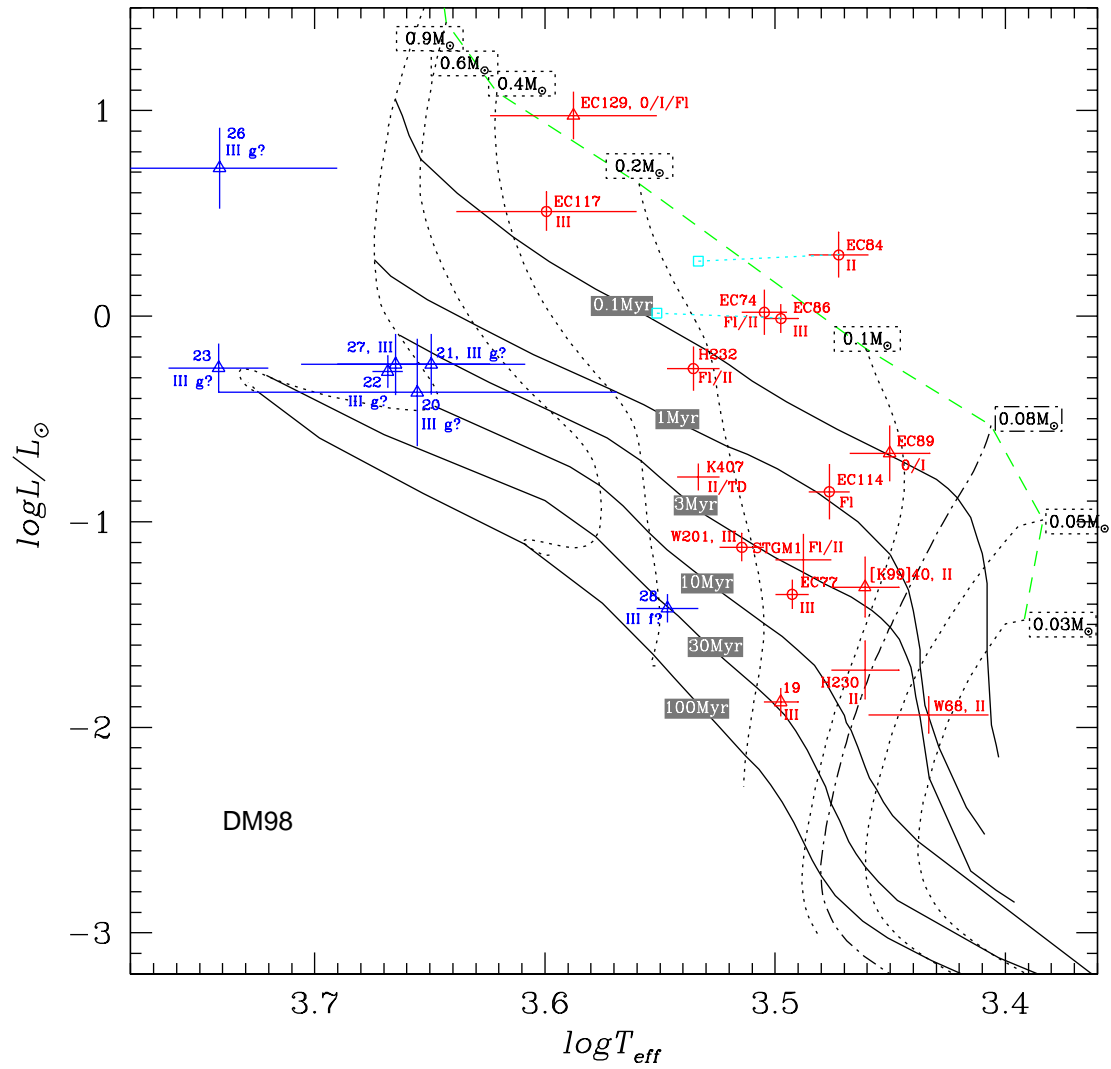}
\caption{Serpens sources overplotted on the evolutionary tracks of \citet{DM97} assuming d$=$380 pc.
Same symbols as in Figures \ref{fig_jhhk} and \ref{hrBCAH}.
}\label{hrDM}
\end{figure}  

\clearpage
\begin{figure} 
\epsscale{0.8}
\plotone{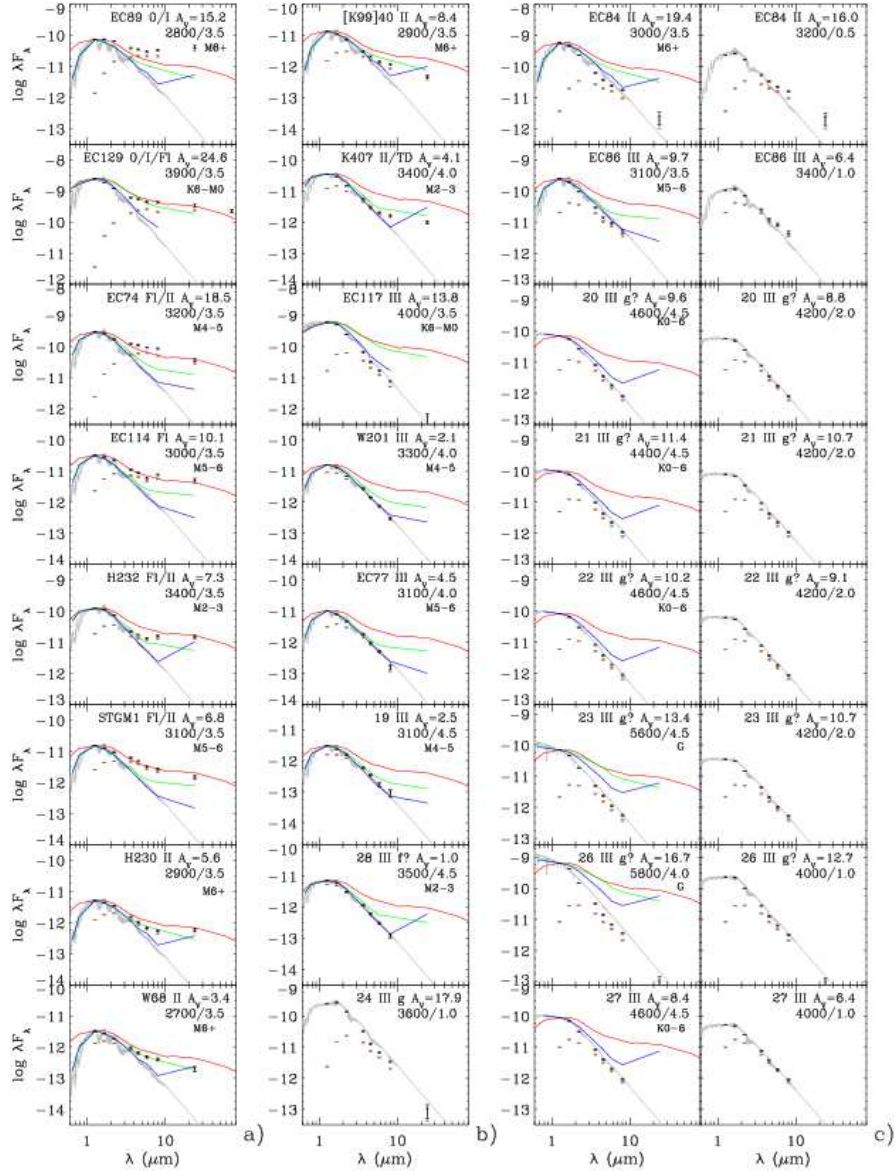}
\caption{SEDs of Serpens sources plotted against the median SED
of CTTSs in Taurus-Auriga ({\it red line}, from \citet{Dalessio99}),
and the ``thick'' ({\it green}) and ``anemic'' ({\it blue})
SpT-dependant disks in IC348 (from \citet{Lada06}, with SpT specified),
scaled to match de-reddened $J$-band
fluxes of Serpens sources.
Note that the median 24 $\mu$m flux for anemic disks
in most cases is based on only 1 -- 2 objects due to sensitivity
limit, and therefore is biased toward the up-turn SEDs.  
{\it Brown dots with errorbars:} observed fluxes, {\it black:}
corrected for extinction.
{\it Gray solid line:} NextGen ($log$ g $>3$) and NextGen-giant ($log$ g $<3$)
photospheric model spectra (also normalized at $J$-band).
$T_{eff}$/$logg$ of the model spectra and $A_{V}$ for de-reddening
are based on our SpTs and the intrinsic colors of YSOs/giants;
for YSO candidates they can be found in Table \ref{tableProp}.
Roman numbers denote {\it Spitzer} classification from the literature
that was done in the absence of the SpT information.}\label{figsed}   
\end{figure}  

\clearpage
\begin{figure} 
%\epsscale{0.9}
\plotone{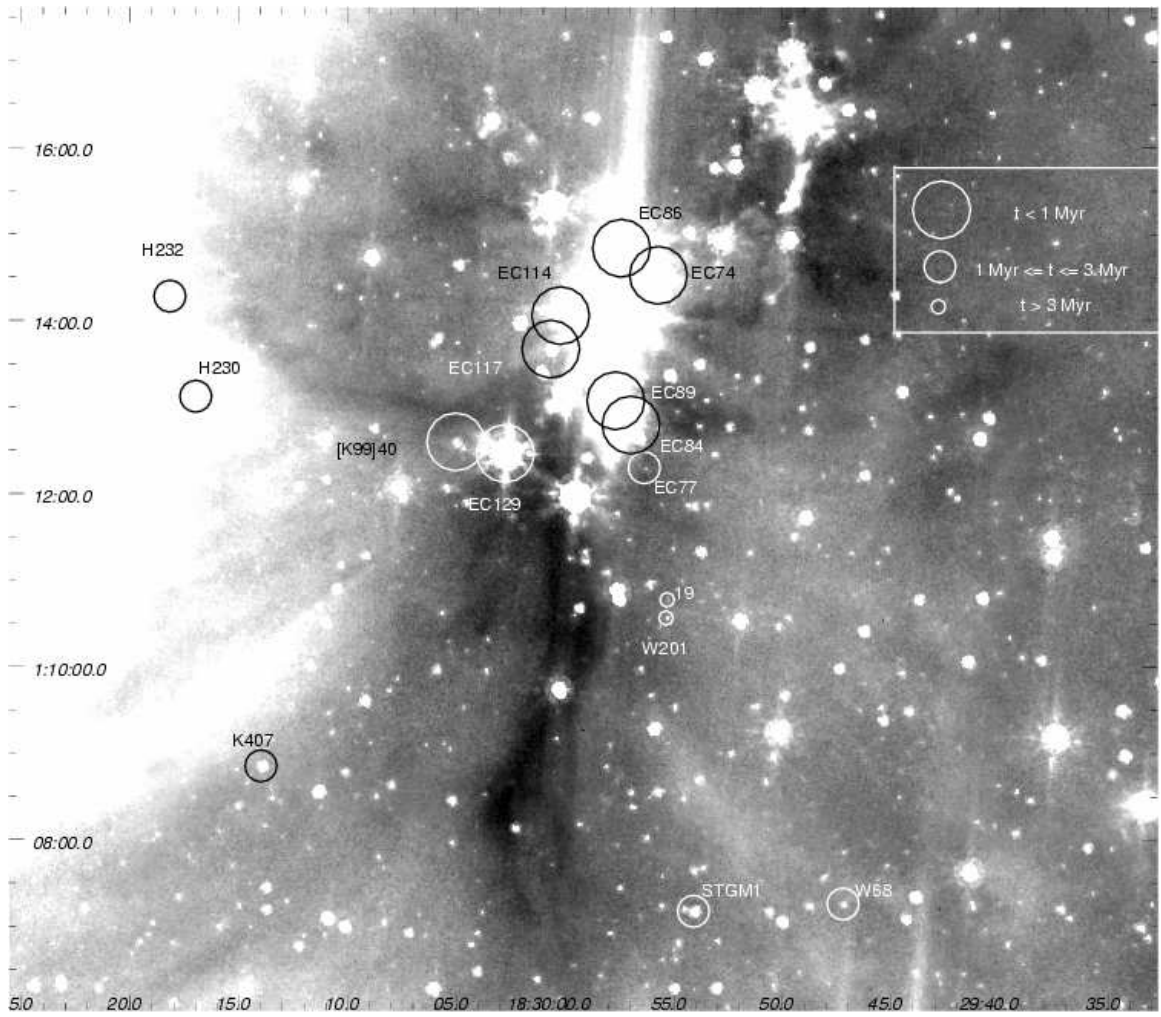}
\caption{YSOs that we could place on the HR diagram
overlaid on the {\it Spitzer} 8 $\mu$m image (from the c2d survey) of the Serpens double core.
The size of the circles designates an age according to the BCAH98 tracks.
The figure illustrates a recent epoch of star formation in the core
at the origin of dark filaments and the older
generation of stars further out.
}\label{figIRACexc}
\end{figure}  

\clearpage
\begin{deluxetable}{rlllrrrrr} 
\tablecolumns{9}  
\tablewidth{0pt} 
\tabletypesize{\tiny}
\tablecaption{Serpens YSO Candidates\label{tableYSO}}
\tablehead{ \colhead{ID}  & \colhead{SED} & Name\tablenotemark{a,b} & \colhead{SpT\tablenotemark{c}} & \colhead{RA(2000)} &
 \colhead{DEC(2000)} & \colhead{J\tablenotemark{c}} &  \colhead{H\tablenotemark{c}} &  \colhead{K\tablenotemark{c}} \\
   &   &   &  & \colhead{hr:min:s} & \colhead{$\degr:\arcmin:\arcsec$} & \colhead{mag} &  \colhead{mag} &   \colhead{mag} }
\startdata 
 1 & 0/I/Fl & SVS2 & $-$ & 18:29:56.89 & +01:14:46.46 & 11.93 $\pm$ 0.02 & 10.57 $\pm$ 0.04 &  9.37 $\pm$ 0.04 \\
 2 & 0/I & EC103   & $-$ & 18:29:58.78 & +01:14:26.09 & 16.75 $\pm$ 0.08 & 14.00 $\pm$ 0.05 & 11.68 $\pm$ 0.03 \\
 3 & 0/I & EC89    & M7.25 $\pm$ 0.75    & 18:29:57.67 & +01:13:04.57 & 16.16 $\pm$ 0.02 & 13.67 $\pm$ 0.02 & 11.93 $\pm$ 0.03 \\
 4 & 0/I/Fl & EC129 & K9 $\pm$3          & 18:30:02.75 & +01:12:27.94 & 15.14 $\pm$ 0.02 & 11.73 $\pm$ 0.02 &  9.93 $\pm$ 0.02 \\
 5 & I/Fl & SVS20A & $-$                 & 18:29:57.73 & +01:14:05.51 & 12.51 $\pm$ 0.02 & 10.05 $\pm$ 0.06 & $-$              \\
 6 & Fl/II & EC74  & M4.5 $\pm$0.5       & 18:29:55.71 & +01:14:31.56 & 15.56 $\pm$ 0.02 & 12.84 $\pm$ 0.02 & 11.22 $\pm$ 0.04 \\
 7 & Fl &  EC114   & M6 $\pm$0.5         & 18:30:00.21 & +01:14:03.57 & 15.59 $\pm$ 0.02 & 13.77 $\pm$ 0.02 & 12.53 $\pm$ 0.04 \\
 8 & Fl/II & H232  & M2.9 $\pm$0.6       & 18:30:18.18 & +01:14:16.79 & 13.34 $\pm$ 0.01 & 11.81 $\pm$ 0.02 & 10.94 $\pm$ 0.03 \\
 9 & Fl/II & STGM1 & M5.4 $\pm$0.6       & 18:29:54.10 & +01:07:10.66 & 15.48 $\pm$ 0.02 & 14.05 $\pm$ 0.02 & 13.13 $\pm$ 0.02 \\
 10 & II & H230    & M6.75 $\pm$0.75     & 18:30:16.99 & +01:13:07.44 & 16.33 $\pm$ 0.02 & 15.01 $\pm$ 0.02 & 14.07 $\pm$ 0.03 \\
 11 & II & $[$K99$]$40 & M6.75 $\pm$0.75 & 18:30:05.03 & +01:12:35.95 & 16.06 $\pm$ 0.02 & 14.54 $\pm$ 0.02 & 13.60 $\pm$ 0.03 \\
 12 & II & W68     & M7.75 $\pm$0.75     & 18:29:47.22 & +01:07:15.75 & 16.21 $\pm$ 0.03 & 15.16 $\pm$ 0.02 & 14.54 $\pm$ 0.02 \\
 13 & II & EC84    & M6.25 $\pm$0.75     & 18:29:56.97 & +01:12:47.76 & 15.11 $\pm$ 0.03 & 12.41 $\pm$ 0.02 & 11.00 $\pm$ 0.03 \\
 14 & II/TD & K407 & M3 $\pm$0.5         & 18:30:13.99 & +01:08:51.16 & 13.80 $\pm$ 0.01 & 12.73 $\pm$ 0.01 & 12.37 $\pm$ 0.02 \\
 15 & III & EC117  & K8.25 $\pm$3.25     & 18:30:00.66 & +01:13:40.35 & 13.44 $\pm$ 0.01 & 11.26 $\pm$ 0.02 & 10.32 $\pm$ 0.03 \\
 16 & III & EC86   & M4.9 $\pm$0.4       & 18:29:57.41 & +01:14:50.34 & 13.26 $\pm$ 0.02 & 11.61 $\pm$ 0.03 & 10.91 $\pm$ 0.04 \\
 17 & III & W201   & M4 $\pm$0.5         & 18:29:55.36 & +01:10:33.92 & 14.11 $\pm$ 0.02 & 13.27 $\pm$ 0.02 & 12.95 $\pm$ 0.03 \\
 18 & III & EC77   & M5.1 $\pm$0.4       & 18:29:56.36 & +01:12:18.08 & 15.26 $\pm$ 0.03 & 14.17 $\pm$ 0.02 & 13.62 $\pm$ 0.03 \\
 19 & III & 19     & M4.9 $\pm$0.4       & 18:29:55.31 & +01:10:46.57 & 16.02 $\pm$ 0.02 & 15.16 $\pm$ 0.02 & 14.63 $\pm$ 0.04 \\ 
\enddata 
%\tablenotetext{a}{Adopted for this paper}
%\tablenotetext{b}{CoKu: \citet{Cohen79}; EC: \citet{Eiroa92}; G: {\it Chandra} ID of \citet{Giardino07};  GCNM: \citet{Giov98}; 
% H: {\it Spitzer} ID of \citet{Harvey07b}; [K99]: \citet{Kaas99} (Table 5); K: {\it ISO} ID of \citet{Kaas04};
% NIRS, STGM: \citet{Sogawa97} (Tables 1 and 2 respectively);  SVS: \citet{Strom76};      
% W: {\it Spitzer} ID of \citet{Winston07}.}
%\tablenotetext{c}{This work.}
%\tablecomments{This table lists FLAMINGOS spectroscopic targets that are likely YSOs in the cloud.} 
\end{deluxetable}

\tablenum{1}
\begin{deluxetable}{rllcr} 
\tablecolumns{5}  
\tablewidth{0pt} 
\tabletypesize{\tiny}
\tablecaption{(cont.)}
\tablehead{ \colhead{ID} & Name\tablenotemark{a,b} & \colhead{Other Names\tablenotemark{b}} &   c2d Name\tablenotemark{d}    & \colhead{Mask-Slit} }
\startdata 
 1   & SVS2 & EC82 NIRS4 STGM22  CoKu-Ser/G7                             & J182956.9+011447 & m2-02\\
     &      &                    GCNM87 K307 H176 W9 G43                 &                  &      \\
 2   & EC103 & STGM20 GCNM112 K326 H190 W4                                  & J182958.8+011426 & m2-03\\
 3   & EC89 & STGM13 GCNM97 H181 W12                                        & J182957.7+011306 & m2-08\\
 4   & EC129 & NIRS10 STGM10 GCNM160 K347                                & J183002.7+011228 & m3-10\\
     &   &      H208 W10                                                 &                  &      \\
 5   & SVS20A & EC90 NIRS5 STGM18 GCNM98                                   & J182957.7+011406 & m2-04\\
     &   &    K314 H182 W35 G48                                          &                  &      \\
 6   & EC74 & STGM21 GCNM76 K298 H171                                     & J182955.7+011432 & m3-03\\
     &   &  W38 G37                                                      &                  &      \\
 7   &  EC114 & STGM17 GCNM131 H200 W28 G62                                  & J183000.2+011404 & m3-05\\
 8   & H232 & W40 G83                                                     & J183018.2+011417 & m1-05\\
 9   & STGM1 & K287 H164 W39                                              & J182954.1+010711 & m2-34\\
 10  & H230 & W76                                                           & J183017.0+011308 & m1-07\\
 11  & $[$K99$]$40 & EC152 H216 W54                                         & J183005.0+011236 & m2-10\\
 12  & W68 & -                                                              & J182947.2+010716 & m4-33\\
 13  & EC84 & NIRS3 STGM11 GCNM90 K309                                      & J182957.0+011248 & m3-09\\
     &  &  W85 G44                                                      &                  &      \\
 14  & K407 & H229 W166                                                  & J183014.0+010852 & m1-25\\
 15  & EC117 & NIRS9 GCNM135 K338 W216 G65                                 & J183000.6+011340 & m3-06\\
 16  & EC86 & GCNM93 W190 G45                                              & J182957.4+011450 & m3-02\\
 17  & W201 & G36                                                          & J182955.4+011034 & m3-17\\
 18  & EC77 & STGM9 GCNM80 W204 G39                                        & J182956.4+011218 & m2-11\\
 19  & 19 & -                                                              & J182955.3+011047 & m2-18\\ 
\enddata 
\tablenotetext{a}{Adopted for this paper}
\tablenotetext{b}{CoKu: \citet{Cohen79}; EC: \citet{Eiroa92}; G: {\it Chandra} ID of \citet{Giardino07};  GCNM: \citet{Giov98}; 
 H: {\it Spitzer} ID of \citet{Harvey07b}; [K99]: \citet{Kaas99} (Table 5); K: {\it ISO} ID of \citet{Kaas04};
 NIRS, STGM: \citet{Sogawa97} (Tables 1 and 2 respectively);  SVS: \citet{Strom76};      
 W: {\it Spitzer} ID of \citet{Winston07}.}
\tablenotetext{c}{This work.}
\tablenotetext{d}{Cores to Disks {\it Spitzer} Legacy catalogue.}
\tablecomments{This table lists FLAMINGOS spectroscopic targets that are likely YSOs in the cloud.} 
\end{deluxetable}

\clearpage
\tablenum{2}
\begin{deluxetable}{rllrrrrrcr} 
%\rotate
\tablecolumns{10}  
\tablewidth{0pt} 
\tabletypesize{\tiny}
\tablecaption{Serpens Stars of Unknown Status\label{tableField}}
\tablehead{ \colhead{ID}        & \colhead{SED}  & \colhead{SpT\tablenotemark{a}} & \colhead{RA(2000)} & \colhead{DEC(2000)} &
\colhead{J\tablenotemark{b}}  & \colhead{H\tablenotemark{b}} & \colhead{K\tablenotemark{b}}   & c2d Name\tablenotemark{c} & \colhead{Mask-}\\
                                & \colhead{   }  & \colhead{   }     & \colhead{hr:min:s} & \colhead{$\degr:\arcmin:\arcsec$} &
\colhead{mag}  & \colhead{mag} & \colhead{mag}  & & \colhead{Slit \#}  
}
\startdata
 20 & III & K4.5 $\pm$ 6.5 / g? & 18:29:53.89 & +01:08:42.01 & 14.62 $\pm$   0.02  &  12.96 $\pm$  0.02  & 12.31 $\pm$  0.02   & J182953.9+010842 & m2-27 \\
 21 & III & K4.5 $\pm$ 2.5 / g? & 18:30:09.24 & +01:11:17.66 & 14.76 $\pm$   0.02  &  12.91 $\pm$  0.02  & 12.13 $\pm$  0.03   & J183009.2+011118 & m1-14 \\
 22 & III & K3.5 $\pm$ 0.5 / g? & 18:30:11.04 & +01:12:10.95 & 14.57 $\pm$   0.02  &  12.89 $\pm$  0.02  & 12.22 $\pm$  0.03   & J183011.0+011211 & m1-10 \\
 23 & III & G7 $\pm$ 3 / g? & 18:29:57.54 & +01:07:26.97     & 15.69 $\pm$   0.02  &  13.82 $\pm$  0.01  & 13.08 $\pm$  0.02   & J182957.5+010727 & m2-33  \\
 24 & III & K-M g & 18:29:47.44 & +01:09:59.16               & 15.60 $\pm$   0.02  &  12.73 $\pm$  0.02  & 11.40 $\pm$  0.02   & J182947.4+010959 & m4-23 \\
 25 & III & M g & 18:29:55.94 & +01:09:17.49                 & 12.96 $\pm$   0.01  &  10.80 $\pm$  0.02  & 10.00 $\pm$  0.02   & J182955.9+010918 & m3-21 \\
 26 & III & G5 $\pm$ 7 / g? & 18:29:53.83 & +01:11:20.13     & 14.21 $\pm$   0.03  &  12.02 $\pm$  0.03  & 11.20 $\pm$  0.03   & J182953.8+011120 & m3-14  \\
 27 & III & K3.5 $\pm$ 2.5 & 18:30:11.07 & +01:10:30.78      & 14.03 $\pm$   0.02  &  12.54 $\pm$  0.01  & 12.01 $\pm$  0.03   & J183011.1+011031 & m1-17 \\
 28 & III & M2.25 $\pm0.75$ f? & 18:29:56.21 & +01:10:20.13  & 14.69 $\pm$   0.02  &  13.93 $\pm$  0.02  & 13.69 $\pm$  0.03   & J182956.2+011020 & m2-20 \\
 29 & III & A & 18:30:12.33 & +01:06:46.68                   & 12.59 $\pm$   0.01  &  11.75 $\pm$  0.02  & 11.49 $\pm$  0.02   & J183012.3+010647 & m1-33 \\
 30 & III & A & 18:29:53.70 & +01:07:40.76                   & 13.81 $\pm$   0.03  &  12.77 $\pm$  0.03  & 12.30 $\pm$  0.01   & J182953.7+010741 & m2-32 \\
\enddata 
\tablenotetext{a}{This work. Sources marked as g? are likely field giants; the spectral classification is given for them assuming alternative YSO nature. Source 28 marked as f? is likely a foreground dwarf.}
\tablenotetext{b}{This work.}
\tablenotetext{c}{Cores to Disks {\it Spitzer} Legacy catalogue.}
\tablecomments{This table lists FLAMINGOS spectroscopic targets that are likely field stars.}
\end{deluxetable}  

%\clearpage
\tablenum{3}
\begin{deluxetable}{lllcrrccc} 
%\rotate
\tablecolumns{9}  
\tablewidth{0pt} 
\tabletypesize{\tiny}
\tablecaption{Stellar Properties of YSO Candidates from FLAMINGOS Spectra\label{tableProp}}
\tablehead{ \colhead{Name}  & \colhead{SED}   & \colhead{SpT}       & \colhead{$logT_{eff}$\tablenotemark{a}}          & \colhead{$A_{V sp.}$}  &
$logL/L_{\sun}$\tablenotemark{b}  & \colhead{$M/M_{\sun}$}            & \colhead{$M/M_{\sun}$}             & $logg$                 \\
\colhead{    }  & \colhead{  }       & \colhead{   }       & \colhead{          }            & \colhead{           }  &
                                  & \colhead{BCAH98}                  & \colhead{DM98}                     & BCAH98
}
\startdata
   EC89 & 0/I    & M7.25 $\pm$ 0.75        & 3.450 $\pm$ 0.017 & 15.24 $\pm$ 1.4  & -0.67 $\pm$  0.14  &  0.08 $^{+\,0.04}_{-\,0.04}$   & 0.11 & 2.8	  \\
  EC129 & 0/I/Fl & K9 $\pm$3               & 3.588 $\pm$ 0.036 & 24.62 $\pm$ 1.1  &  0.98 $\pm$  0.12  &  1.08 $^{+\,0.32}_{-\,0.31}$	& 0.29 & 2.8	  \\
   EC74 & Fl/II & M4.5 $\pm$0.5           & 3.505 $\pm$ 0.010 & 18.50 $\pm$ 1.0  &  0.02 $\pm$  0.11  &  0.35 $^{+\,0.05}_{-\,0.05}$	& 0.16 & 2.9	  \\
  EC114 & Fl    & M6 $\pm$0.5             & 3.476 $\pm$ 0.009 & 10.08 $\pm$ 1.1  & -0.85 $\pm$  0.13  &  0.13 $^{+\,0.03}_{-\,0.03}$	& 0.14 & 3.3	  \\
   H232 & Fl/II & M2.9 $\pm$0.6           & 3.536 $\pm$ 0.011 &  7.27 $\pm$ 1.0  & -0.25 $\pm$  0.11  &  0.50 $^{+\,0.10}_{-\,0.10}$	& 0.23 & 3.5	  \\
  STGM1 & Fl/II & M5.4 $\pm$0.6           & 3.488 $\pm$ 0.012 &  6.80 $\pm$ 1.0  & -1.19 $\pm$  0.13  &  0.15 $^{+\,0.04}_{-\,0.05}$	& 0.16 & 3.7	  \\
   H230 & II     & M6.75 $\pm$0.75         & 3.461 $\pm$ 0.015 &  5.60 $\pm$ 1.2  & -1.72 $\pm$  0.14  &  0.06 $^{+\,0.02}_{-\,0.02}$	& 0.09 & 3.7	  \\
 $[$K99$]$40 & II& M6.75 $\pm$0.75         & 3.461 $\pm$ 0.015 &  8.40 $\pm$ 1.2  & -1.32 $\pm$  0.15  &  0.07 $^{+\,0.03}_{-\,0.02}$   & 0.11 & 3.4	  \\
    W68 & II     & M7.75 $\pm$0.75         & 3.433 $\pm$ 0.026 &  3.44 $\pm$ 0.8  & -1.94 $\pm$  0.09  &  0.03 $^{+\,0.03}_{-\,0.02}$   & 0.04 & 3.6	  \\
   EC84 & II     & M6.25 $\pm$0.75         & 3.472 $\pm$ 0.013 & 19.42 $\pm$ 0.6  &  0.30 $\pm$  0.11  &  0.23 $^{+\,0.07}_{-\,0.05}$	& 0.10 & 2.3	  \\
   K407 & II/TD  & M3   $\pm$0.5           & 3.533 $\pm$ 0.009 &  4.11 $\pm$ 0.5  & -0.78 $\pm$  0.06  &  0.38 $^{+\,0.08}_{-\,0.05}$	& 0.29 & 3.9	  \\
  EC117 & III    & K8.25 $\pm$3.25         & 3.599 $\pm$ 0.039 & 13.75 $\pm$ 0.6  &  0.51 $\pm$  0.09  &  1.20 $^{+\,0.40}_{-\,0.40}$	& 0.35 & 3.3	  \\
   EC86 & III    & M4.9 $\pm$0.4           & 3.497 $\pm$ 0.007 &  9.65 $\pm$ 0.6  & -0.01 $\pm$  0.07  &  0.30 $^{+\,0.03}_{-\,0.04}$	& 0.15 & 2.9	  \\
   W201 & III    & M4 $\pm$0.5             & 3.514 $\pm$ 0.010 &  2.10 $\pm$ 0.5  & -1.12 $\pm$  0.07  &  0.25 $^{+\,0.05}_{-\,0.04}$	& 0.23 & 4.0	  \\
   EC77 & III    & M5.1 $\pm$0.4           & 3.493 $\pm$ 0.007 &  4.50 $\pm$ 0.6  & -1.35 $\pm$  0.07  &  0.15 $^{+\,0.03}_{-\,0.03}$	& 0.18 & 3.9	  \\
     19\tablenotemark{c} & III    & M4.9 $\pm$0.4           & 3.497 $\pm$ 0.007 &  2.45 $\pm$ 0.5  & -1.88 $\pm$  0.07  &  0.13 $^{+\,0.03}_{-\,0.03}$	& 0.17 & 4.3	  \\
\enddata
\tablenotetext{a}{Using YSO scale of \citet{Luhman03}.}
\tablenotetext{b}{Assuming distance 380 pc.}
\tablenotetext{c}{YSO status is only suggested based on the location on the BCAH98 HR diagram.}
%\tablecomments{}
\end{deluxetable}  

\end{document}